\documentclass[aps,prl,twocolumn,nofootinbib,groupedaddress]{revtex4-1}
\usepackage{amsmath,amsfonts,amssymb,bbm,epsfig}
\UseRawInputEncoding

\def\be{\begin{equation}}
\def\ee{\end{equation}}
\def\bq{\begin{equation}}
\def\eq{\end{equation}}
\def\bqa{\begin{eqnarray}}
\def\eqa{\end{eqnarray}}
\def\roughly#1{\mathrel{\raise.3ex
\hbox{$#1$\kern-.75em\lower1ex\hbox{$\sim$}}}}
\def\lsim{\roughly<}
\def\gsim{\roughly>}

\renewcommand{\theequation}{\arabic{section}.\arabic{equation}}

\begin{document}


\title{The Proton Gluon Distribution from \\
the Color Dipole Picture}



\author{G.R. Boroun}
\affiliation{Department of Physics, Razi University,\\
Kermanshah 67149, Iran}

\author{M. Kuroda}
\affiliation{Center for Liberal Arts, Meijigakuin University\\
Yokohama, Japan}

\author{Dieter Schildknecht}
\affiliation{Fakult\"at f\"ur Physik, Universit\"at Bielefeld\\
D-33501 Bielefeld, Germany\\
and\\
Max-Planck-Institut f\"ur Physik (Werner-Heisenberg-Institut)\\
F\"ohringer Ring 6, D-80805 M\"unchen, Germany}


\date{\today}

\begin{abstract}


Employing the representation of the experimental data on deep inelastic 
electron-proton scattering (DIS) in the color-dipole picture (CDP), we
determine the gluon distribution of the proton at small Bjorken $x$. At
sufficiently large momentum transfer, $Q^2$, the extracted gluon distribution
fulfills the standard evolution equation for the proton structure function. For low
values of $Q^2$, e.g. for $Q^2 = 1.9 {\rm GeV}^2$, the evolution equation for
the proton structure function is violated. The standard procedure of adopting a low-$Q^2$ starting
scale for the extraction of the gluon density is questionable and requires further
investigations.

\end{abstract}


\maketitle


\section{\label{Introduction}I. Introduction}
\renewcommand{\theequation}{\arabic{section}.\arabic{equation}}
\setcounter{section}{1}\setcounter{equation}{0}
The extraction of the gluon-distribution function \cite{Harland, Dulat,
Ball, Buckley} of the proton
from deep-inelastic electron-proton scattering (DIS) \cite{Abramowicz}
rests on the
determination of the fit parameters in an ad-hoc parametrization of
the gluon x-distribution at a low-$Q^2$ input scale.\footnote{In standard
notation, the virtuality of the photon is denoted by $Q^2$, and $x$ is the
Bjorken scaling variable, $x \cong Q^2/W^2$, where $W$ denotes the
virtual-photon-proton center of mass energy.} It is well-known, and
was recently emphasized in ref.\cite{Pelicer} that the results on the gluon
distribution function from different
collaborations \cite{Harland, Dulat, Ball, Buckley}
show a significant spread \footnote{See Fig.2 in ref.[6]} in the low-x,
low-$Q^2$ domain. The quality of the fits is considered [6] not to be satisfactory.
This led to the introduction of phenomenological power corrections
\cite{HM,Cooper} to the
structure functions, and to modifications \cite{Pelicer} of the DGLAP evolution
equations \cite{Lipatov} for
the gluon distribution by a non-linear term \cite{Gribov, Mueller}.

In the color-dipole picture (CDP) [12], [13],[14], [15]
 of DIS,
the photoabsorption cross section
at low $x$ is represented in terms of the quark-antiquark proton
($q \bar q p$) cross section, $\sigma_{(q \bar q) p} (W^2)$. The
corresponding forward scattering amplitude is given by an ansatz for
the color-gauge-invariant interaction of $q \bar q$ states with the
proton via two-gluon exchange. The $q \bar q$ states form a massive
continuum as a function of the $q \bar q$ masses (as observed in
$e^+ e^- \to q \bar q$ annihilation), including a smooth interpolation
of the low-lying vector meson peaks. Accordingly, the massive $q \bar q$
continuum starts at a mass squared of $m^2_0 \lsim m^2_\rho$, where
$m^2_{\rho^0}$ denotes the square of the $\rho^0$ meson mass. The general structure
(at leading order of $\alpha_s (Q^2)$) of the two-gluon exchange interaction
of $\gamma^*g \to q \bar q$ (where $\gamma^*$ denotes the photon of virtuality
$Q^2$) from the pQCD improved parton model is assumed to remain valid when
proceeding from large values of $Q^2$ (the genuine region of pQCD) to small
values of $Q^2$, for $Q^2$ towards $Q^2 = 0$. In the CDP, accordingly, the
representation of DIS is continued to include the $Q^2$ towards $Q^2 = 0$
limit, and the photoproduction cross section $\sigma_{\gamma p}
(W^2)$ may be used as a normalization of the $\gamma^*p$ interaction cross
section, $\sigma_{\gamma^* p} (W^2, Q^2)$.

The basic quantity of the CDP ansatz, the cross section $\sigma_{(q \bar q)p}
(W^2)$, is dependent on $W^2$ via $\Lambda^2_{sat} (W^2) \equiv C_1
\left( \frac{W^2}{1 {\rm GeV}^2} \right)^{C_2}$, where $C_1$ and $C_2$
are adjustable parameters.

The dominance of the $\gamma^*g \to q \bar q$ interaction is common
to both, the description of DIS at low $x$ in the CDP and in the pQCD-based
parton model, where instead of $\sigma_{(q \bar q)p} (W^2)$ the gluon
distribution function $G(x,Q^2)$ enters as the basic quantity. The basic
mechanism of $\gamma^*g \to q \bar q$ being the same, the underlying
gluon distribution may be deduced from the CDP representation.

In the present paper, at leading order in the strong coupling
$\alpha_s (Q^2)$, we accordingly determine the gluon distribution,
$G(x, Q^2)$, from the CDP representation  of the DIS experimental
data.

Adopting an implicit validity of the $\gamma^*g \to q \bar q$ interaction,
a gluon distribution may also be extracted from a fit to the experimental
data not manifestly being based on the $\gamma^*g \to q \bar q$
interaction of the CDP. Assuming implicit validity of the $\gamma^*g
\to q \bar q$ interaction, we in addition accordingly deduce
$G (x, Q^2)$ from the Froissart-bounded representation  of the DIS data.
We find results consistent with the ones based on the CDP. Our results on the gluon distribution  do not exclusively depend on the CDP representation of the DIS experimental data.  

In the present paper, accordingly, the leading order gluon distribution of the perturbative-QCD-improved (pQCD) parton model is
extracted from the representation of the proton structure functions $F_2 (x, Q^2)$
and $F_L (x, Q^2)$ in the CDP. We find that at low $x$, for $Q^2$ sufficiently large ($Q^2{\gtrsim}10~ {\rm GeV}^2$ to $20~ {\rm GeV}^2$), the experimental results on DIS parametrised by the CDP fulfill evolution in distinction from low values of $Q^2$ (e.g. for $Q^2=1.9~ {\rm GeV}^2$), where due to hadronlike behaviour of DIS,  the standard evoluation equation is strongly violated. This result suggests that the unsatisfactory fits mentioned in Ref.[6]\footnote{See Fig.2 in ref.[6]} at low $Q^2$ are due to an improper use of a low-$Q^2$ input scale, such as $Q^2=1.9~ {\rm GeV}^2$, in the widely employed analysis of the DIS experimental results.

In Section II, we briefly review the determination of the gluon distribution at
leading order of $\alpha_s (Q^2)$ from the pQCD improved parton model. In Section III,
we review the relevant parts of the CDP. In Section IV, as an alternative to the CDP, we consider and summarise the
Froissart-bounded representation of $F_2 (x, Q^2)$. In Section V, the results obtained
for the gluon distribution are presented for both the CDP and the Froissart-bounded parametrization of the DIS data. At low $Q^2~~(Q^2 = 1.9 {\rm GeV}^2)$, we
observe drastic differences between our results on the gluon distribution and the results published by several different collaborations. Note that for the extraction of the gluon distribution in Sections II to V, no use is being made of the pQCD
evolution equations \cite{Lipatov}.

In Sections VI and VII, we examine the consistency between the low-$Q^2$ dependence at
low-$x$ of our results for the gluon distribution and the validity of the evolution
equation for the structure function $F_2 (x, Q^2)$. At large $Q^2$, specifically for
approximately $20~{\rm GeV}^2 \lsim Q^2 \lsim 100 {\rm GeV}^2$, the CDP gluon distribution
is consistent with the validity of evolution, leading to the important constraint of
$C_2 = 0.29$ for the exponent in the dependence of $F_2 (x, Q^2)$ on $(W^2)^{C_2} =
(Q^2 / x)^{C_2}$. The predicted value of $C_2 \cong 0.29$ agrees with
experiment. At low-$Q^2$, the logarithmic derivative of the structure function
$F_2 (x, Q^2)$ deviates from the prediction of the evolution equation by a factor of magnitude up to 3. Final Conclusions are
presented in Section VIII.

\section{II. The determination of the gluon distribution function in pQCD.}
\renewcommand{\theequation}{\arabic{section}.\arabic{equation}}
\setcounter{section}{2}\setcounter{equation}{0}
In the pQCD improved parton model, the longitudinal structure function $F_L (x, Q^2)$ to
lowest order of $\alpha_s (Q^2)$ is given by \cite{Martin}
\be
F_L (x, Q^2) = \frac{\alpha_s(Q^2)}{4 \pi} \left( \frac{16}{3} I_F + 8 \sum Q^2_q I_g\right),
\label{eq:6.1}
\ee
where
\be
I_F \equiv I_F (x, Q^2) = \int^1_x \frac{dy}{y} \left( \frac{x}{y} \right)^2 F_2 (y, Q^2),
\label{eq:6.2}
\ee
and
\be
I_g \equiv I_g (x, Q^2) = \int^1_x \frac{dy}{y} \left( \frac{x}{y} \right)^2 \left(
1 - \frac{x}{y} \right) G (x, Q^2),
\label{eq:6.3}
\ee
the gluon density being denoted by $G(x,Q^2)=xg(x,Q^2)$, and $\sum Q^2_q = 10/9$ for four flavors of quarks.
The term proportional to $I_F$ in (2.1) is due to $\gamma^*$-quark (antiquark)
interaction with gluon emission,
\be
\gamma^* + q (\bar q) \to q (\bar q) + g,
\label{eq:6.4}
\ee
while the term proportional to $I_g$ in (2.1) originates from quark-antiquark
production,
\be
\gamma^* + g \to q \bar q.
\label{eq:6.5}
\ee

The successful representation of the experimental data on DIS in the CDP is based on
incorporating \cite{Cvetic} 
the $q \bar q$-color-dipole interaction (2.5)
with the gluon field of the proton
\bqa
 (\gamma^* \to q \bar q) + g  \to & q \bar q\nonumber\\
\mathrm{or}~~~~~~~~~~~~~~~~\nonumber\\
\gamma^* +(g \to q \bar q) \to & q \bar q,
\label{eq:6.4}
\eqa
into the mass-dispersion relation of
Generalized Vector
Dominance (GVD) \cite{Sakurai}. In the CDP  the photon makes a transition  to  $q \bar q$ states that interact with gluons, or, equivalently, the photon (exclusively) intracts with the $q \bar q$ sea originating from  $g \to q \bar q$ transitions, see (2.6).

 Accordingly, motivated by the CDP representation of the experimental data, we assume that the gluon
term associated with
reaction (2.5) is the dominant one in the pQCD expression (2.1),
\bqa
F_L (x,Q^2)= \frac{2 \alpha_s (Q^2)}{\pi} \sum_q Q^2_q I_g(x,Q^2). 
\eqa
In the CDP of low-$x$ DIS, the large-$Q^2$ pQCD equality (2.1)  is replaced by (2.7), thus excluding (2.4). 
 
For a wide range of different gluon distributions, independently of their
specific form, the evaluation of $I_{g}(x,Q^2)$ from (2.3) in (2.7) yields the well-known
approximate result [19, 14 ] for the longitudinal
structure function (2.7)
\be
F_L \left(\xi_L x, Q^2\right) = \frac{\alpha_s(Q^2)}{3 \pi} \sum_q Q^2_q
G (x,Q^2).
\label{eq:2.4}
\ee
A given gluon distribution determines the longitudinal proton structure
function at a rescaled value of $x \to \xi_L x$.
The rescaling factor $\xi_L$ in (\ref{eq:2.4}) has the preferred value of
$\xi_L \cong 0.40$ \cite{Martin}.

We have explicitly tested the validity of the replacement of the
pQCD integral representation (2.7) between the longitudinal structure
function of the CDP and the gluon distribution by the simple
proportionality (2.8) for the determination of the gluon distribution.
Determining the gluon distribution from (2.8) by inserting the
explicit CDP expression for $F_L \left(\xi_L x, Q^2\right)$ given by
(5.1) below, and inserting the result into (2.7), we indeed reproduce
the CDP input for $F_L (x, Q^2)$ within a deviation of up to at
most 3.5 \%. Compare the results in Table 1.

\begin{table}[h]
	\caption{The table shows the ratio of $F_L (W^2,Q²)$ from (2.7)
		upon insertion of the gluon distribution from the proportionality
		(2.8), to the CDP input for $F_L (x, Q^2)$ from (5.1). The ratio
		is indeed close to unity.}
	\begin{tabular}{|l|c|c|c|}
		\hline
		& $W^2 = 10^5$ & $W^2 = 10^4$ & $W^2 = 10^3$ \\
		& $[ {\rm GeV}^2 ] $ & $[ {\rm GeV}^2 ] $ & $[ {\rm GeV}^2 ] $ \\
		\hline
		$Q^2 = 100$ & 1.043 & 1.037 & 0.981 \\
		$[ {\rm GeV}^2 ]$ & & & \\
		\hline
		$Q^2 = 50$ & 1.041 & 1.037 & 1.012 \\
		$[ {\rm GeV}^2 ]$ & & & \\
		\hline
		$Q^2 = 10$ & 1.035 & 1.035 & 1.030 \\
		$[ {\rm GeV}^2 ]$ & & & \\
		\hline
		$Q^2 = 2$ & 1.021 & 1.026 & 1.029 \\
		$[ {\rm GeV}^2 ]$ & & & \\
		\hline
		$Q^2 = 1$ & 1.015 & 1.021 & 1.026 \\
		$[ {\rm GeV}^2 ]$ & & & \\
		\hline
	\end{tabular}
\end{table}

The simple proportionality (2.8) indeed consistently provides a gluon
distribution that fulfills the pQCD relationship (2.7). The $Q^2$ dependence of the gluon distribution in (2.7) is indeed determined by the $Q^2$ dependence of the proton
structure function of the CDP in (5.1) upon insertion into (2.8).




\section{III. The Color-Dipole Representation.}
\renewcommand{\theequation}{\arabic{section}.\arabic{equation}}
\setcounter{section}{3}\setcounter{equation}{0}

For a detailed representaion of the CDP, we refer to Refs.[14] and [15]. In a brief summary, the photon interaction (2.5)  with the gluon field in the proton at low values of $x$ is interpreted as fluctuation of the photon into $q\bar{q}$-color-dipole states, see (2.6), that interact with the gluon field in the proton via color-gauge-invariant gluon couplings to quark and antiquark states.

 In Fig. 1\footnote{The photoabsorption cross section (\ref{eq:3.1}) and
(\ref{eq:3.2}) in good approximation is determined by $\eta (W^2,Q^2)$, i.e.
$\sigma_{\gamma^*p} (W^2,Q^2) \cong \sigma_{\rho^*p} (\eta (W^2,Q^2))$ except
for a weak (logarithmic) $W$-dependence due to $\sigma^{(\infty)} (W^2)$ in
(\ref{eq:3.1}) or $\sigma_{\gamma p} (W^2)$ in (\ref{eq:3.2}). The parameter
$\xi$ (not to be confused with $\xi_L$) is fixed at $\xi = \xi_0 = 130$,
and for the range of $\eta (W^2,Q^2)$ of relevance for the present paper,
one may put $\xi \to \infty$.}, reproduced from refs. \cite{Ku-Schi, Kuroda}, we show  the results for the photoabsorption cross
section, $\sigma_{\gamma^*p} (W^2,Q^2)$, in the CDP. The results are obtained
from the explicit analytic expression 

\begin{figure}[h]
\includegraphics[width=8cm]{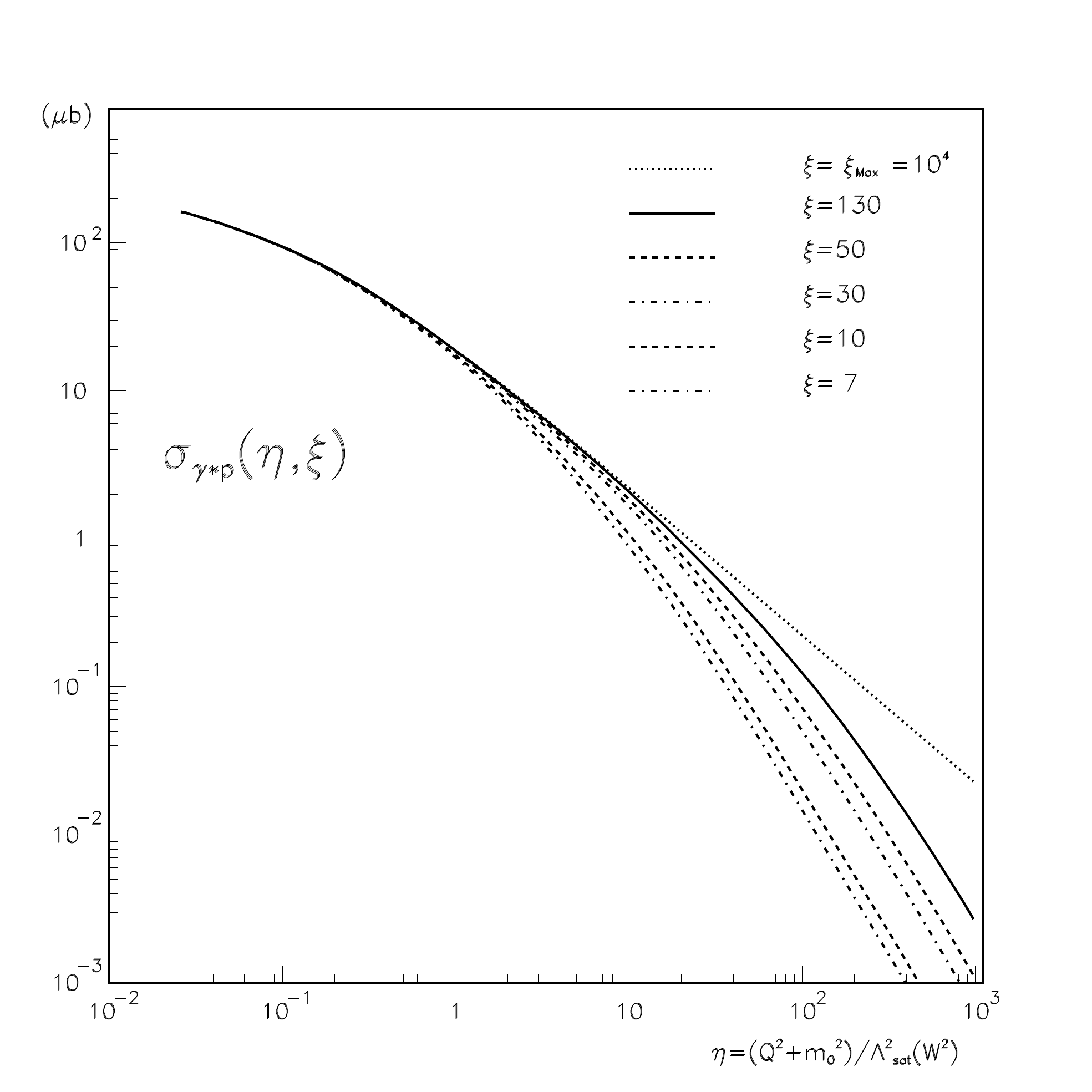}%
\caption{\label{Fig1}The theoretical results for the photoabsorption cross section
$\sigma_{\gamma^*p} (\eta (W^2,Q^2), \xi)$ in the CDP as a function of
the low-x scaling variable $\eta (W^2,Q^2) = (Q^2 + m^2_0)/\Lambda^2_{sat}
(W^2)$ for different values of the parameter $\xi$ that determines the
(squared) mass range $M^2_{q \bar q} \le m^2_1 (W^2) = \xi \Lambda^2_{sat}
(W^2)$ of the $\gamma^* \to q \bar q$ fluctuations that are taken into
account. The experimental results   for 
 $\sigma_{\gamma^*p}(\eta(W^2,Q^2),\xi)$ 
 lie on the full line corresponding
to $\xi = \xi_0 = 130$, compare refs. \cite{Ku-Schi, Kuroda}.
}
\end{figure}
for $\sigma_{\gamma^*p} (W^2,Q^2)$,
\bqa
\sigma_{\gamma^*p} (W^2,Q^2) & = & \sigma_{\gamma^*_Lp} (W^2,Q^2) +
\sigma_{\gamma^*_Tp} (W^2,Q^2)  \nonumber \\
& = & \frac{\alpha R_{e^+e^-}}{3 \pi} 
\sigma^{(\infty)} (W^2)  \nonumber \\
&&\times\left( I^{(1)}_T \left( \frac{\eta}{\rho}, \frac{\mu}{\rho} \right)
G_T (u) \right. \nonumber \\
&&~~~~~\left.+ I^{(1)}_L
(\eta, \mu) G_L (u)
\right),\nonumber\\
&&
\label{eq:3.1}
\eqa
derived from an ansatz \cite{Ku-Schi,Kuroda,Schild} for the W-dependent dipole cross 
section that essentially, via the structure of the
coupling of the quark-antiquark state to two gluons, represents the
color-gauge-invariant interaction of the $q \bar q$ dipole with the
gluon-field in the nucleon. 
In (\ref{eq:3.1}), $R_{e^+e^-}=3\sum_qQ_q^2$, where $q$ runs over the active quark
flavors, and $Q_q$ denotes the quark charge.
The smooth transition to $Q^2 = 0$
photoproduction in (\ref{eq:3.1}) allows one \cite{Kuroda}  to replace 
(see (2.37) in ref. \cite{Kuroda})
$\sigma^{(\infty)} (W^2)$, which stems from the normalization of the
$q \bar q$-dipole-proton cross section\footnote{Note that the photon fluctuates [13,16] into on-mean-shell massive $q \bar q$ states (see e.g. Ref.[18] for an explicit proof of this point) implying a dipole-proton cross section that depends on $W^2$  (besides the transverse dipole size)  and not on $x$, a dependence on $x$ nevertheless being adopted frequently without  justification.}, by the photoproduction cross section, and
(\ref{eq:3.1}) becomes
\bqa
\sigma_{\gamma^*p} (W^2, Q^2) && = \frac{\sigma_{\gamma p} (W^2)}
{\lim\limits_{\eta \to
\mu (W^2)} I_T^{(1)} 
\left( \frac{\eta}{\rho},
\frac{\mu (W^2)}{\rho}\right) G_T (u)}  \nonumber \\
&& \hspace*{-1.4cm}\times\left( I^{(1)}_T  \left(
\frac{\eta}{\rho}, \frac{\mu}{\rho} \right) G_T (u) + I^{(1)}_L (\eta, \mu)
G_L (u) \right). ~~
\label{eq:3.2}
\eqa
We note that $I_L^{(1)} (\eta, \mu)$ vanishes in the photoproduction
$Q^2 = 0$ limit of $\eta(W^2,Q^2=0)=m_0^2/\Lambda^2_{sat}(W^2) \equiv \mu(W^2)$,
and  $G_T (u\equiv{\xi\over\eta}) \simeq 1$,
and for later reference we also note
\be
\lim_{\eta \to \mu (W^2)} I^{(1)}_T \left( \frac{\eta}{\rho},
\frac{\mu (W^2)}{\rho} \right) = \ln \frac{\rho}{\mu (W^2)}.
\label{eq:3.3}
\ee
For the general explicit analytic expressions for the functions $I^{(1)}_T
\left( \frac{\eta}{\rho},\frac{\mu (W^2)}{\rho} \right)$ and
$I^{(1)}_L (\eta, \mu)$  we refer to Appendix A. The functions
$G_T \left( u\equiv\frac{\xi}{\eta}
\right)$ and $G_L \left( u\equiv\frac{\xi}{\eta} \right)$ are given by
\be
G_T (u) =  \frac{2u^3 + 3 u^2 + 3u}{2 (1+u)^3} \simeq
\left\{ \begin{array}{l@{\quad,\quad}l}
\frac{3}{2} \frac{\xi}{\eta} & (\eta \gg \xi), \\
1 - \frac{3}{2} \frac{\eta}{\xi} & (\eta \ll \xi),
\end{array} \right.
\label{eq:3.4}
\ee
and
\be
G_L (u) = \frac{2u^3 + 6u^2}{2 (1+u)^3} \simeq
\left\{ \begin{array}{l@{\quad,\quad}l}
3 \left( \frac{\xi}{\eta}\right)^2 & (\eta \gg \xi),\\
1 - 3 \left( \frac{\eta}{\xi} \right)^2 & (\eta \ll \xi),
\end{array} \right.
\label{eq:3.5}
\ee
where $u \equiv \xi / \eta$, and the constant parameter $\xi$ restricts
the masses of the contributing mass $q \bar q$ states via
\be
M^2_{qq} \le m^2_1 (W^2) = \xi \Lambda^2_{sat} (W^2).
\label{eq:3.6}
\ee

The numerical results for the photoabsorption cross section
in Fig. 1 are obtained by numerical evaluation of (\ref{eq:3.2})
upon insertion of a $\left( \ln W^2 \right)^2$ fit to the experimental results for
the photoproduction cross section $\sigma_{\gamma p} (W^2)$ from
the Particle Data Group \cite{PDG}. 
The results in Fig. 1 were
obtained for $W = 275 ~{\rm GeV}$ from
\bqa
\sigma_{\gamma p} (W^2) = & 0.003056 \left( 34.71 + \frac{0.3894\pi}{M^2} \ln^2 \frac{W^2}{M^2_p + M)^2}
\right) \nonumber \\
& + 0.0128 \left( \frac{(M_p + M)^2}{W^2} \right)^{0.462}.
\label{eq:3.7}
\eqa
In (\ref{eq:3.7}), $M_p$ denotes the proton mass, $M = 2.15~{\rm GeV}$ and $\sigma_{\gamma\rho}
(W^2)$ is given in units of millibarn.

Before going into more detail, we note
that the full curve in Fig. 1, which  for the parameter $\xi$ corresponds 
to the choice of
$\xi = \xi_0 = 130$,  provides
a representation of the full set of experimental data on 
$\sigma_{\gamma^*p} (W^2,Q^2)$ at low $x \cong Q^2/W^2$, compare Fig. 8 in ref. \cite{Ku-Schi}.

In (\ref{eq:3.1}) and (\ref{eq:3.2}), the low-x scaling variable $\eta (W^2,Q^2)$ is
given [13] by
\be
\eta \equiv \eta (W^2,Q^2) = \frac{Q^2 + m^2_0}{\Lambda^2_{sat} (W^2)},
\label{eq:3.8}
\ee
with
\be
\mu \equiv \mu(W^2) = \eta (W^2,Q^2 = 0) = \frac{m^2_0}{\Lambda^2_{sat}
(W^2)},
\label{eq:3.9}
\ee
the "saturation scale"\footnote{The value of $\Lambda^2_{sat} (W^2)$ via $\eta(W^2,Q^2){\cong}~ 5$ determines the transition from color transparency of $\eta(W^2,Q^2){ \gtrsim}~ 5$ to hadronlike saturation of  $\eta{\lesssim}5$. }, $\Lambda^2_{sat} (W^2)$, being parametrized by
\be
\Lambda^2_{sat} (W^2) = C_1 \left( \frac{W^2}{1 {\rm GeV}^2} \right)^{\LARGE{C_2}},
\label{eq:3.10}
\ee
and numerically the results in Fig.1 are based on
\bqa
m^2_0 & = & 0.15~ {\rm GeV}^2, \nonumber \\
C_1  & = & 0.31~ {\rm GeV}^2;~~~C_2 = 0.27.
\label{eq:3.11}
\eqa
The parameter $\rho = const$ in (\ref{eq:3.1}) and (\ref{eq:3.2})
is related to the longitudinal-to-transverse
ratio $R (W^2,Q^2)$ of the photoabsorption cross section, and
approximately we have $R(W^2,Q^2) \simeq 1/2 \rho$ for
$\eta (W^2,Q^2) \gg \mu (W^2)$, while $R(W^2,Q^2) = 0$ for $Q^2 = 0$.
The total cross section $\sigma_{\gamma^*p} (W^2,Q^2)$ is fairly insensitive 
to the value of $\rho$ for realistic values of $\rho$ around $\rho \cong 1$, and  
the evaluation presented in Fig. 1 is based on 
$\rho = {4\over 3}$.

In terms of the photoabsorption cross sections $\sigma_{\gamma^*_{L,T} p}
(W^2,Q^2)$, in (\ref{eq:3.1}) and (\ref{eq:3.2}), the proton structure
functions, relevant for the extraction of the gluon-density distribution,
see (\ref{eq:2.4}), are given by
\be
F_{L,T} (W^2, Q^2) = \frac{Q^2}{4 \pi^2 \alpha} \sigma_{\gamma^*_{L,T} p}
(W^2, Q^2)
\label{eq:3.12}
\ee
and
\bqa
& F_2 (W^2, Q^2) = \frac{Q^2}{4 \pi^2 \alpha} \sigma_{\gamma^*p} (W^2, Q^2) \nonumber \\
& = \frac{Q^2}{4 \pi^2 \alpha} \left( \sigma_{\gamma^*_T p} (W^2, Q^2) +
\sigma_{\gamma^*_L p} (W^2, Q^2)\right).
\label{eq:3.13}
\eqa
Upon introducing the longitudinal-to-transverse ratio $R(W^2,Q^2)$, the longitudinal structure function becomes
\be
F_L (W^2,Q^2) = \frac{R}{1+R} F_2 (W^2,Q^2).
\label{eq:3.14}
\ee
Explicitly, according to (\ref{eq:3.2}), we have
\be
R(W^2,Q^2) = \frac{I_L^{(1)} (\eta, \mu) G_L(u)}{I_T^{(1)} 
\left( \frac{\eta}{\rho}, \frac{\mu}{\rho}\right) G_T (u)}.
\label{eq:3.15}
\ee

For $Q^2=0$, as a consequence of electromagnetic gauge invariance,
\be
R(W^2,Q^2 = 0) = 0,
\label{eq:3.16}
\ee
while for $\eta(W^2,Q^2)$, restricted by 
the interval of $1 \ll \eta (W^2,Q^2) \ll \xi = \xi_0 = 130$
that will be of relevance subsequently,
we have
\be
R (W^2,Q^2) \approx \frac{1}{2 \rho},
\label{eq:3.17}
\ee
with $\rho = const$ in the vicinity of $\rho \cong 1$,
and, accordingly, (\ref{eq:3.14}) becomes
\bqa
F_L (W^2,Q^2) && = \frac{1}{2 \rho + 1} F_2 (W^2,Q^2), \nonumber \\
&& (1 \ll \eta (W^2,Q^2) \ll \xi_0).
\label{eq:3.18}
\eqa
While (\ref{eq:3.18}) holds strictly for sufficiently large $Q^2$, 
even for low values of $Q^2{\gtrsim}1 \textrm{GeV}^2$ it may be used
as an approximation of (3.14). The accuracy of (3.18) as approximation for low $Q^2$ is seen by comparing the results in Fig.3 and Fig.4 in Section V below.

For large $Q^2$, specifically for $10 {\rm GeV}^2 \lsim Q^2 \lsim 100 {\rm GeV}^2$
the range of $Q^2$ of particular relevance in connection with pQCD,
the experimental data on the proton structure function $F_2 (W^2,Q^2)$ in
(\ref{eq:3.13}) depend on the single variable $W^2$, compare Fig. 2 quoted from
ref. \cite{Ku-Schi}.

\begin{figure}[h]
\includegraphics[width=8cm]{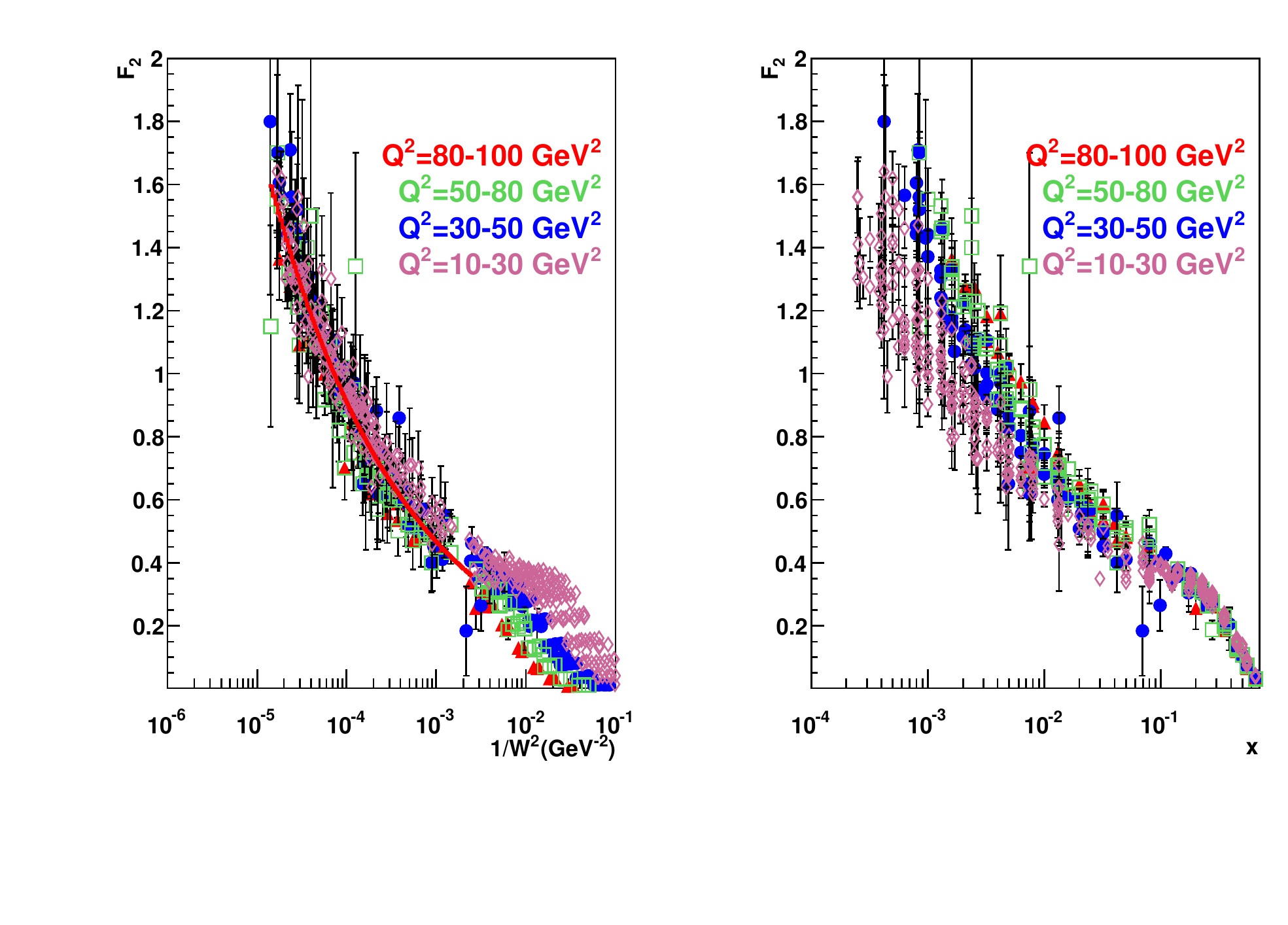}%
\vspace*{-1cm}
\hspace{5cm}{\bf 2a}\hspace{4cm}{\bf 2b}
\caption{\label{Fig2}In Fig.2a we show the experimental data for $F_2(x \cong
Q^2/W^2, Q^2)$ as a function
of $1/W^2$, and in Fig.2b, for comparison, as a function of $x$. The theoretical prediction
based on (\ref{eq:3.19}) with (\ref{eq:3.20}) is also shown in Fig.2a.
}
\end{figure}

A simple two-parameter eye-ball fit to the experimental data in Fig. 2 yields
\cite{Ku-Schi}
\be
F_2 (W^2) = f_2 \left( \frac{W^2}{1 {\rm GeV}^2} \right)^{C_2},
\label{eq:3.19}
\ee
where \footnote {The value of the exponent $C_2=0.29$ is consistent  \cite{D21} with the independent fit to the DIS experimental data based on the hard-Pomeron conjecture leading to the exponent $C_0 \cong 0.30 \pm 0.1.$ }
\be
f_2 = 0.063,~~C_2 = 0.29.
\label{eq:3.20}
\ee

The fit result is understood as a consequence of the CDP in the large-$\eta(W^2,Q^2)$
approximation. The structure function $F_2 (W^2,Q^2)$ for $Q^2 \gg \Lambda^2_{sat} (W^2)$
according to (\ref{eq:3.13}) upon substitution of (\ref{eq:3.1}) takes the simple form
\bqa
F_2 (W^2,Q^2) &=& \frac{R_{e^+e^-} \sigma^{(\infty)} (W^2)}{24 \pi^3}
\frac{1+2 \rho}{3} \Lambda^2_{sat} (W^2) \nonumber \\
& \times & \left(1 + 0 \left( \frac{1}{\eta} \right) \right).
\label{eq:3.21}
\eqa
Upon specifying $\Lambda^2_{sat} (W^2) = C_1 \left( \frac{W^2}{1 {\rm GeV}^2} \right)^{C_2}$
according to (\ref{eq:3.10}), the structure function (\ref{eq:3.21}) can directly be
compared with (\ref{eq:3.19}),
\bqa
F_2 (W^2,Q^2) && = \frac{R_{e^+e^-} \sigma^{(\infty)} (W^2)}{24 \pi^3}
\frac{1+2 \rho}{3} C_1 \left( \frac{W^2}{1 {\rm GeV}^2} \right)^{C_2} \nonumber \\
&& \times \left(
1 + 0 \left( \frac{1}{\eta} \right) \right)  \label{eq:3.22} \\
&& \equiv \hat{f}_2 (W^2) \left( \frac{W^2}{1 {\rm GeV}^2} \right)^{C_2} 
\left( 1 + 0 \left( \frac{1}{\eta} \right) \right). \nonumber
\eqa
The product $R_{e^+e^-} \sigma^{(\infty)} (W^2)$ is determined by the photoproduction
cross section (\ref{eq:3.7}) according to
\be
R_{e^+e^-} \sigma^{(\infty)} (W^2) = \frac{3 \pi}{\alpha}
\frac{\sigma_{\gamma p} (W^2)}{\ln \frac{\rho \Lambda^2_{sat} (W²)}{m^2_0}},
\label{eq:3.23}
\ee
compare to (\ref{eq:3.1}) to (\ref{eq:3.3}). 

For definiteness, in Table 2,
we explicitly present the input for the evaluation of $F_2$ 
according to (\ref{eq:3.22}) that leads to
\be
\hat{f}_2 (W^2) = \left\{ \begin{array}{l@{\quad,\quad}l}
0.067 & {\rm for}~W^2 = 10^4 {\rm GeV^2}, \\
0.068 & {\rm for}~W^2 = 10^5 {\rm GeV^2}.
\end{array} \right.
\label{eq:3.24}
\ee

\begin{table}[h]
\caption{The evaluation of $\hat{f}_2 (W^2)$ defined by (\ref{eq:3.22}). The
parameters $\rho$ and $m^2_0$ in (\ref{eq:3.22}) and (\ref{eq:3.23}) are given
by $\rho = 4/3$ and $m^2_0 = 0.15 {\rm GeV}^2$.}
\begin{tabular}{|l|c|c|}
\hline
 $W^2~[{\rm GeV}^2]$ & $10^4$ & $10^5$ \\
 \hline
 $\sigma_{\gamma p}~[mb] $ & 0.146 & 0.175 \\
 \hline
 $\Lambda^2_{sat} (W^2) = C_1 \left( \frac{W^2}{1 {\rm GeV}^2} \right)^{C_2}$ & ~ & ~ \\
 $C_1 = 0.31~ {\rm GeV}^2$, & 4.48 & 8.74 \\
 $C_2 = 0.29$	& ~ & ~ \\
 \hline
 $R_{e^+e^-} \sigma^{(\infty)} (W^2)~[mb]$ & 50.9 & 51.9  \\
 \hline
 $\hat{f}_2 (W^2)$ & 0.067 & 0.068  \\
 \hline
\end{tabular}
\end{table}
These values of $\hat{f}_2 (W^2)$ are approximately 6 \% to 7 \% larger than the
values from the eye-ball fit (\ref{eq:3.19}) in (\ref{eq:3.20}). The structure function
$F_2 (W^2, Q^2)$ is given by
\be
F_2 \cong \left\{ \begin{array}{l@{\quad,\quad}l}
0.96 & {\rm for}~W^2 = 10^4~{\rm GeV}^2, \\
1.91 & {\rm for}~W^2 = 10^5~{\rm GeV}^2.
\end{array} \right.
\label{eq:3.25}
\ee

For the determination of the gluon distribution according to (\ref{eq:2.4}),
the structure functions in (\ref{eq:3.12}) and (\ref{eq:3.13}) have to
be evaluated at a rescaled or shifted value of $x \to \xi_L x$,
corresponding to a shift of $W^2 \to \xi^{-1}_L W^2$. In terms of the low-x
scaling variable $\eta (W^2, Q^2)$, the shift becomes
\be
\eta (W^2, Q^2) = \frac{Q^2 + m^2_0}{\Lambda^2_{sat} (W^2)} \to
\frac{Q^2 + m^2_0}{\Lambda^2_{sat} \left(\xi_L^{-1} W^2\right)}.
\label{eq:3.25}
\ee
For
\be
\Lambda^2_{sat} (W^2) = C_1 \left( \frac{W^2}{1~{\rm GeV}^2} \right)^{C_2},
\label{eq:3.26}
\ee
to be employed subsequently, the shift becomes
\be
\eta (W^2, Q^2) \to \xi_L^{C_2} \eta (W^2, Q^2).
\label{eq:3.27}
\ee

The photoabsorption cross section (\ref{eq:3.2}) essentially only
depends on $\eta (W^2, Q²)$,
\be
\sigma_{\gamma^*_{L,T}p} (W^2, Q^2) \cong \sigma_{\gamma^*_{L,T}p}
\left( \eta (W^2, Q^2) \right),
\label{eq:3.28}
\ee
since the $Q^2 = 0$ photoproduction factor in (\ref{eq:3.2}) depends only
weakly on $W^2$. The rescaling shift (\ref{eq:3.27}) applied to the
cross section (\ref{eq:3.2}), accordingly, amounts to
\be
\sigma_{\gamma^*_{L,T}p} \left( \eta (W^2, Q^2)\right) \to
\sigma_{\gamma^*_{L,T}p} \left( \xi_L^{C_2} \eta (W^2,Q^2) \right).
\label{eq:3.29}
\ee
Numerically, for $C_2 = 0.29$ and $\xi_L = 0.4$, we have
\be
\xi_L^{C_2} \eta = 0.4^{0.29} \eta \cong 0.77 \eta (W^2,Q^2).
\label{eq:3.30}
\ee
Explicitly, the gluon distribution (\ref{eq:2.4}) becomes
\be
\alpha_s (Q^2) G(x,Q^2) = \frac{3 \pi}{\sum_q Q^2_q} \frac{Q^2}{4 \pi^2 \alpha}
\sigma_{\gamma^*_Lp} \left( \xi_L^{C_2} \eta (W^2,Q^2) \right).
\label{eq:3.31}
\ee
We note that (3.32) provides the connection between the gluon distribution (the basic quantity of the pQCD- improved parton model) and the dipole cross section (the basic quantity of the CDP), since $\sigma_{\gamma^*_Lp}$ in (3.32) may be replaced by the dipole cross section $\sigma_{(q\bar q) p}(r_{\bot},z(1-z), W^2=\frac{Q^2}{x})$, according to 
\bqa
\sigma_{\gamma^*_Lp}( W^2=\frac{Q^2}{x},Q^2)&=&\int{dz}\int d^2r_{\bot}|\Psi_{L}(r_{\bot},z(1-z),Q^2)|^2 \nonumber\\
&&\hspace*{-0.5cm}\times\sigma_{(q\bar q) p}(r_{\bot},z(1-z), W^2=\frac{Q^2}{x}).
\label{eq:3.33}
\eqa
The empirically verified low-x scaling of the photoabsorption cross section
in the variable $\eta (W^2, Q^2)$ translates from the photoabsorption
cross section to the gluon distribution. For
sufficiently large $Q^2$, with $\sigma_{\gamma^*_Lp} \propto
\Lambda^2_{sat} (W^2)/Q^2$ [15], according to (3.32) we obtain the asymptotic behavior of
\be
\alpha_s (Q^2) G(x,Q^2) \propto \Lambda^2_{sat} (W^2).
\label{eq:3.32}
\ee
The quantity $\Lambda^2_{sat} (W^2)$,
also known as saturation scale, determines the gluon distribution function
in the pQCD large-$Q^2$ limit.

 The arguments in Section II and the present Section III may be summarized as follows: In the CDP, at leading order of pQCD, DIS at low $x$ is (successfully) described as interaction of the photon with the $q \bar q$ sea according to (2.6). This implies that the pQCD equation (2.1)  for the longitudinal proton structure function  can be approximated by the gluon contribution, see (2.7), and, finally the gluon distribution can be determined from (2.8). On the left-hand side, the proton structure function from the CDP has to be substituted.
 
  Independently  from the CDP, the approximation of (2.1) by (2.8) as leading term at large $Q^2$ was justified  [19] by evaluating suitable models for the gluon distribution.

 Note that the pQCD equation (2.8)  for its validity requires sufficiently large values of $Q^2$. Since the CDP description by the proton structure functions to be substituted into (2.8) includes the low-$Q^2$ limit, the use of (2.8) contains a smooth extrapolation of the gluon density to low values of $Q^2$.

\section{IV. The Froissart-bounded Representation of $F_2 (x,Q²)$.}
\renewcommand{\theequation}{\arabic{section}.\arabic{equation}}
\setcounter{section}{4}\setcounter{equation}{0}

A very accurate fit to the measured proton structure function of DIS, due to Block
et al. \cite{Block}, is provided by a representation of the DIS data satisfying the $(\log W^2)^2$
bound derived from general-field-theory principles 
by Froissart \cite{Froissart}.  Even though  the determination of the gluon distribution (2.8) contains the CDP representation for the proton structure function, one may expect that different precise fits to the DIS data will at least approximately lead to the same results for the gluon distribution. In this spirit, we shall employ the Froissart-bounded representaion in addition to the CDP representation for the proton structure function.

The results from DIS at low $x \le 0.1$ and for a large range of $Q^2$
from $0.15 {\rm GeV}^2 \le Q^2 \le 3000 {\rm GeV}^2$ are represented
by the structure function \cite{Block}
\bqa
F_2 (x,Q^2) &=& D(Q^2) (1-x)^n \left[ C(Q^2) \right. \nonumber \\
&+& A (Q^2) \ln \left( \frac{1}{x} \frac{Q^2}{Q^2+\mu^2} \right) \nonumber \\
&+& \left. B (Q^2) \ln^2 \left( \frac{1}{x} \frac{Q^2}{Q^2+\mu^2} \right) \right].
\label{eq:4.1}
\eqa
The power $n$ and the scale $\mu^2$ are given by $n = 11.49 \pm 0.99$ and
$\mu^2 = 2.82 \pm 0.290~ {\rm GeV}^2$.
The logarithmic dependence on $Q^2$ of the functions $A(Q^2), B(Q^2), C(Q^2)$
and $D(Q^2)$ is given in Appendix B and Table B1 reproducing Table II
from ref. \cite{Block}.

In distinction from the CDP, the Froissart-bounded fit does not explicitly 
provide a representation of the longitudinal-to-transverse ratio $R$. When
deducing the gluon distribution function, the value of $R$ from the CDP
will have to be adopted.

\section{V. The Results for the Gluon Distribution Function.}
\renewcommand{\theequation}{\arabic{section}.\arabic{equation}}
\setcounter{section}{5}\setcounter{equation}{0}

We turn to the explicit results for the gluon distribution which follow from
the evaluation of (2.8) upon substitution of the representation of the 
DIS experimental results in the CDP and the Froissart-bounded representation, as
given in Sections III and IV, respectively.

Substitution into (2.8) of the CDP results (\ref{eq:3.12}) with (\ref{eq:3.2})
for $F_L (W^2,Q^2)$ yields
\bqa
&&\hspace*{-0.5cm}\alpha_s (Q^2)G(x,Q^2) = \frac{3 \pi}{\sum_q Q^2_q} F_L \left(\xi_L x,
Q^2\right) =
\frac{9Q^2}{4\pi \alpha R_{e^+e^-}}   \nonumber \\
&&\hspace*{-0.5cm}\times \left( \frac{\sigma_{\gamma p} (W^2)}{\left( \ln \frac{\rho}{\mu} \right)
G_T \left( \frac{\xi}{\eta} \right)} I_L^{(1)} (\eta, \mu) G_L 
\left( \frac{\xi}{\eta} \right) \right)\Bigg |_{W^2 \to \xi_L^{-1}
W^2}. 
\label{eq:5.1}
\eqa
The variables $\eta = \eta (W^2,Q^2)$ and $\mu (W^2)$ are given by (see
(\ref{eq:3.8}), (\ref{eq:3.9})),
\bqa
\eta (W^2,Q^2) &=& \frac{Q^2 + m^2_0}{\Lambda^2_{sat} (W^2)}, \nonumber \\
\mu (W^2) &=& \frac{m^2_0}{\Lambda^2_{sat} (W^2)},
\label{eq:5.2}
\eqa
where
\be
\Lambda^2_{sat} (W^2) = C_1 (W^2)^{C_2} = 0.31 \left( \frac{W^2}{1 {\rm GeV}^2} \right)^{0.29}~{\rm GeV}^2,
\label{eq:5.3}
\ee
with $C_2 = 0.29$ from (\ref{eq:3.20}) and
\be
m^2_0 = 0.15~ {\rm GeV}^2 .
\label{eq:5.4}
\ee
For the assumed  number of four flavors,
\be
R_{e^+e^-} \equiv 3 \sum_q Q^2_q = \frac{10}{3}.
\label{eq:5.5}
\ee
The parameter $\rho$, denoting the size enhancement of transversely relative
to longitudinally polarized $q \bar q$ fluctuations, is given by  [24]
\be
\rho = \frac{4}{3} .
\label{eq:5.6}
\ee
The parameter $\xi$ that stands for the upper limit, $\xi \Lambda^2_{sat}
(W^2)$, of masses of $q \bar q$ fluctuations actively contributing, is given
by $\xi = 130$. At the relevant (low) values of $\eta (W^2,Q^2)$, the limit
of $\xi \to \infty$ and $G_T \left( \frac{\xi}{\eta} \right) = G_L
\left( \frac{\xi}{\eta} \right) = 1$ (see (\ref{eq:3.4}) and (\ref{eq:3.5})) 
may be taken without loss of generality, compare Fig. 1.
The $\log (W^2)$ fit to the photoproduction experimental results is given by 
(\ref{eq:3.7}), and, finally, the shift factor $\xi_L$ according to (\ref{eq:2.4})   
is given by
\be
\xi_L = 0.4 .
\label{eq:5.7}
\ee

The essential effect of the shift $W^2 \to \xi_L^{-1} W^2$, in (\ref{eq:5.1}) amounts
to
\be
\eta (W^2,Q^2) \to \xi_L^{C_2} \eta (W^2,Q^2),
\label{eq:5.8}
\ee
compare (\ref{eq:3.29}), the $Q^2 = 0$ photoproduction term 
in (\ref{eq:5.1}), including its normalization
by the denominator in (\ref{eq:5.1}), being hardly affected due to its weak
dependence on $W^2$.

According to (2.8), the gluon distribution may equivalently be expressed
in terms of the structure function $F_2 \left( \xi_L x, Q^2\right)$ instead of
$F_L \left( \xi_L x, Q^2 \right)$, together with
the longitudinal-to-transverse ratio $R$. For sufficiently large $Q^2$, according
to (2.8) and (\ref{eq:3.18}), with $R = 1/2 \rho$,
\be 
\alpha_s (Q^2) G (x,Q^2) = \frac{9\pi}{R_{e^+e^-}} \frac{1}{2\rho + 1} F_2
\left( \xi_L x, Q^2 \right).
\label{eq:5.10} 
\ee
The representation (\ref{eq:5.10}), even
at small $Q^2$ may be used as an approximation of the gluon distribution. 

For $Q^2$ sufficiently large, specifically for $20 ~{\rm GeV}^2 \lsim Q^2 \lsim 100~ {\rm GeV}^2$,
the large-$Q^2$ limit of (\ref{eq:5.1}) given by (\ref{eq:3.19}) becomes
relevant. It implies the simple large-$Q^2$ representation of
\bqa 
&& \alpha_s (Q^2) G (x,Q^2) = \frac{9 \pi}{R_{e^+e^-}} \frac{1}{(2 \rho + 1)} F_2 (\xi_L x, Q^2)
\nonumber \\
&=& \frac{9 \pi}{R_{e^+e^-}} \frac{1}{(2 \rho + 1)} f_2 \xi_L^{-0.29} \left(
\frac{W^2}{1 {\rm GeV}^2} \right)^{C_2 = 0.29},
\label{eq:5.9} 
\eqa
where $f_2 = 0.063$, and $\rho = \frac{4}{3}$.

\begin{figure}[h]
\includegraphics[width=8cm]{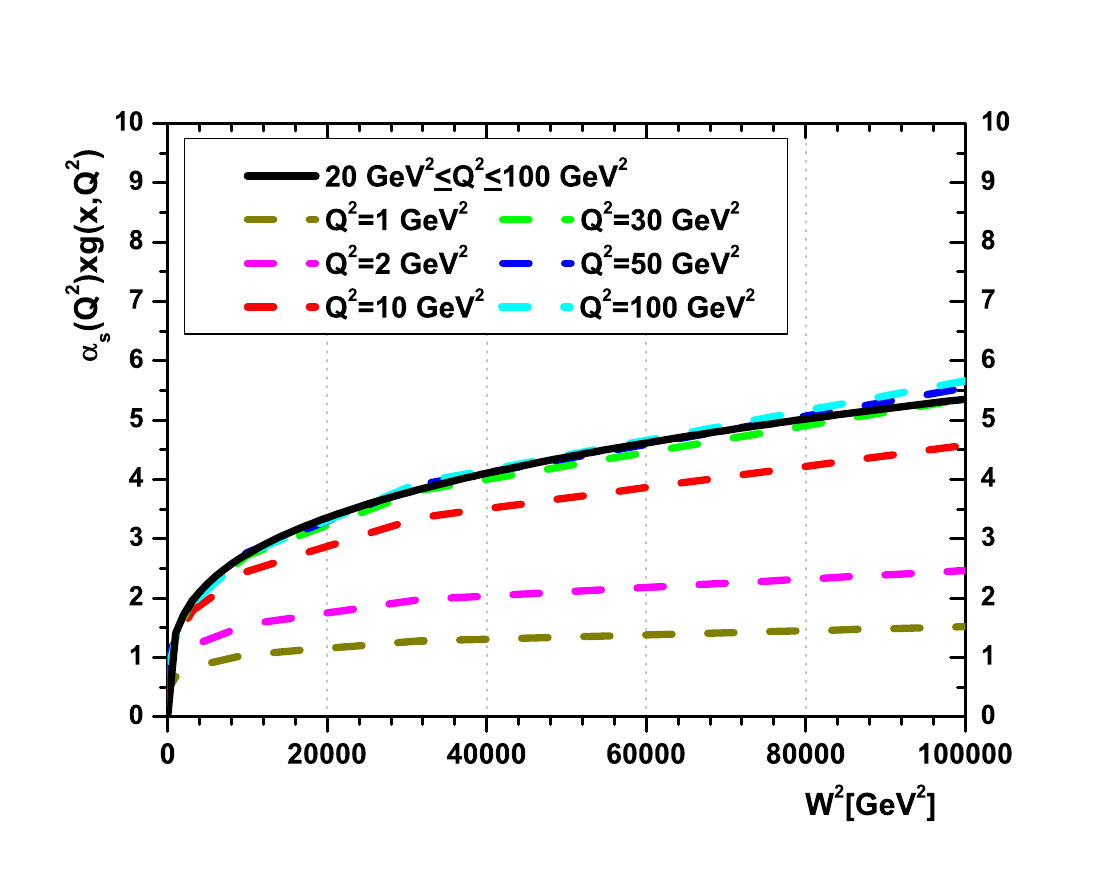}%
\caption{\label{Fig3}The gluon distribution $\alpha_s (Q^2) x g(x,Q^2) \equiv \alpha_s
 (Q^2) G (x,Q^2)$
of the CDP, compare (\ref{eq:5.1}), as a function of $W^2$ for various values of
$Q^2$. The solid line shows the asymptotic limit (\ref{eq:5.9}) that is reached
at $Q^2 \gsim 30 {\rm GeV}^2$.
}
\end{figure}

In Fig. 3, we show the gluon distribution (2.8) deduced from $F_L (W^2, Q^2)$ according to
(\ref{eq:5.1}) as a function of
$W^2$ for various values of $Q^2$, where $1 {\rm GeV}^2 \le Q^2 \le 100 {\rm GeV}^2$.
The results in Fig. 3, for $Q^2$ sufficiently above $Q^2 \cong 10 {\rm ~GeV}^2$,
indeed converge towards the asymptotic representation (\ref{eq:5.9}).

\begin{figure}[h]
\includegraphics[width=8cm]{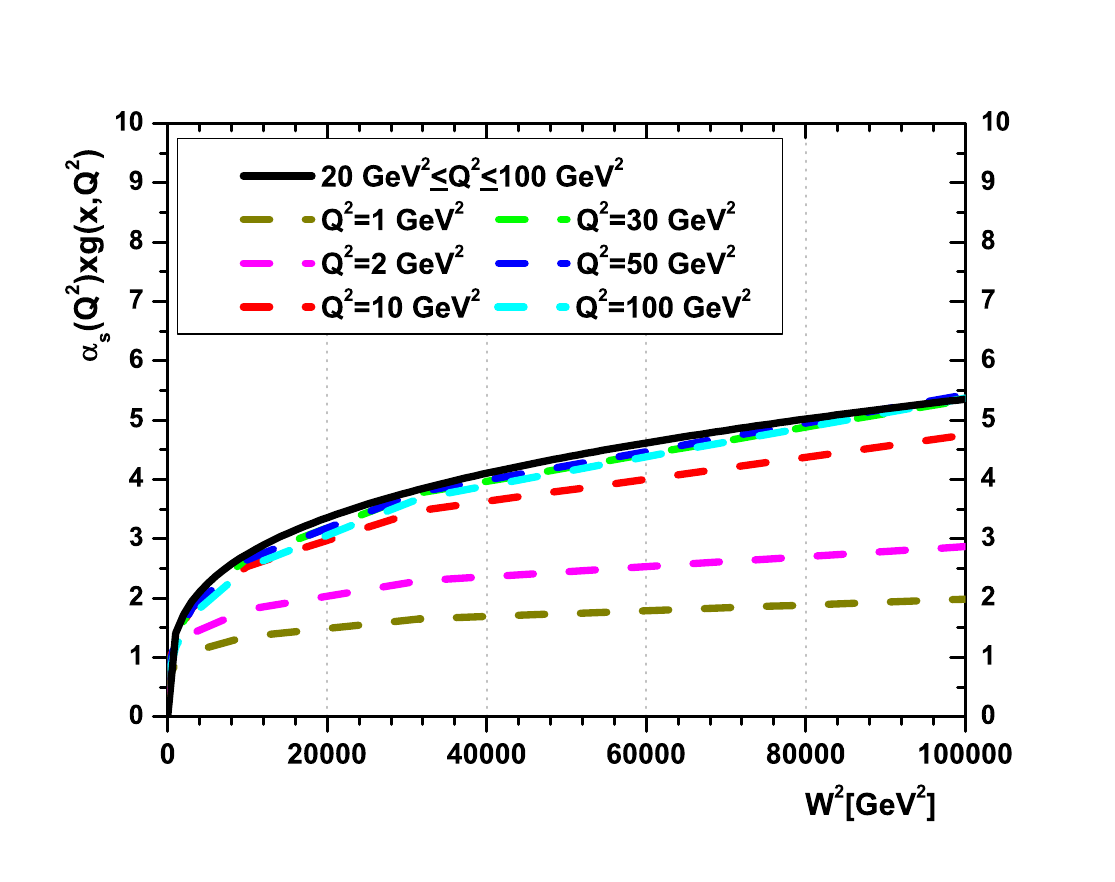}%
\caption{\label{Fig4}As Fig. 3, but based on (\ref{eq:5.10}), employing $R = const =
1/2 \rho = 3/8$. Compare text for details.
}
\end{figure}

The
results in Fig. 4, for low $Q^2$, $1 {\rm GeV}^2 \le Q^2 \le 10 {\rm GeV}^2$,  compared with the
results in Fig. 3, show the enhanced gluon distribution resulting
from employing the large-$Q^2$ approximation (\ref{eq:5.10}). For sufficiently large $Q^2$, the
results in Fig. 4, based on $F_2 \left( \xi_L x, Q^2 \right)$ according to (3.18) and (\ref{eq:5.10}), 
coincide with the ones in Fig. 3
based on $F_L \left( \xi_L x, Q^2 \right)$ according to (\ref{eq:5.1}).

\begin{figure}[h]
\includegraphics[width=8cm]{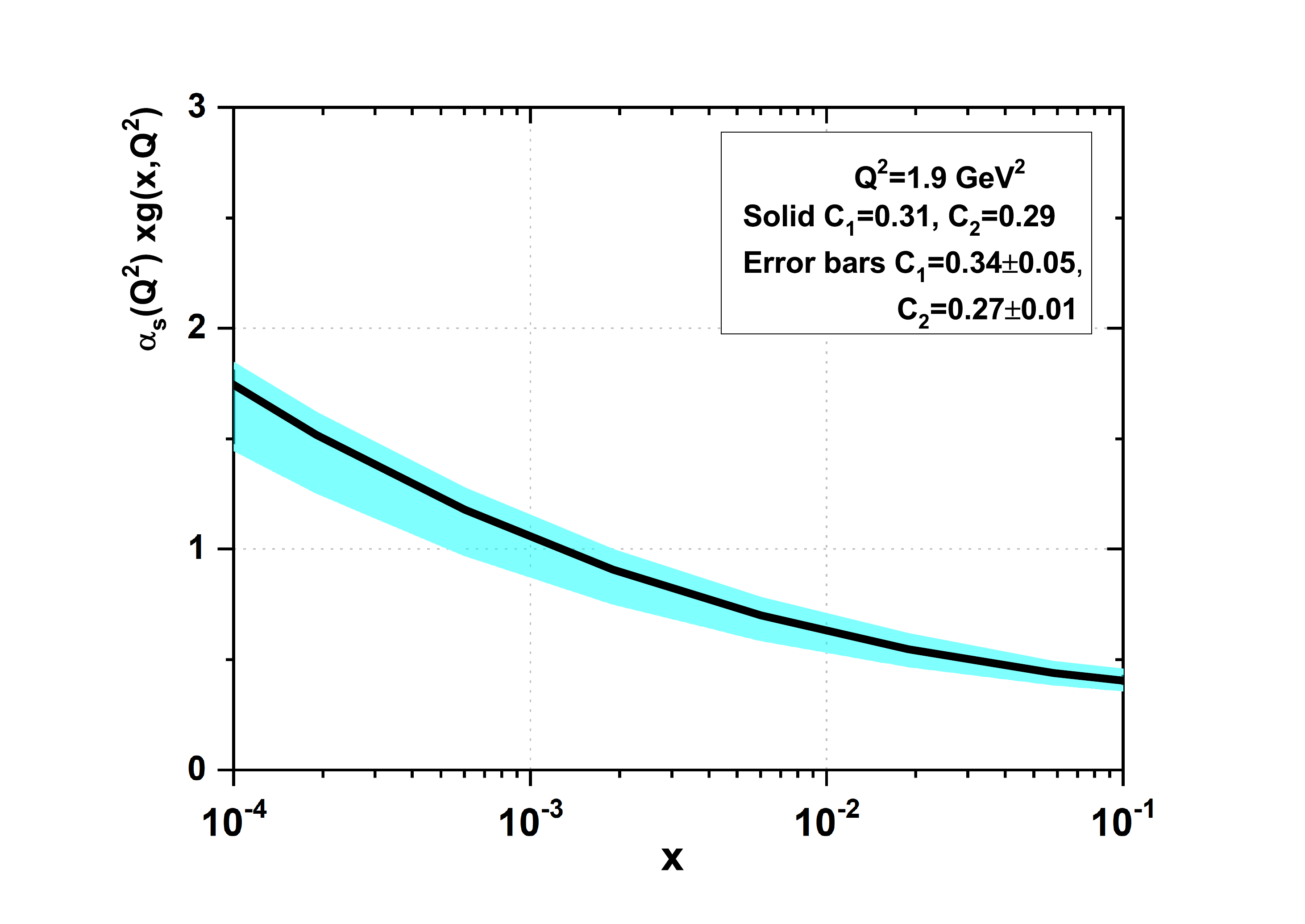}%
\caption{\label{Fig5}The gluon-distribution function $\alpha_s (Q^2) 
xg (x,Q^2)$ of the CDP as a function of $x \cong Q^2/W^2$ at the (low) value of
$Q^2 = 1.9~ {\rm GeV}^2$, the scale frequently used as input scale \cite{Pelicer}.
The solid curve is due to  $C_{1}=0.31$ and $C_{2}=0.29$ and the uncertainties are due to  $C_{1}=0.34{\pm}0.05$ and 
$C_{2}=0.27{\pm}0.01$ \cite{Schild} (G. Cvetic, D. Schildknecht, B. Surrow, M. Tentyukov, Eur. Phys. J. C {\bf 20}, 77 (2001)).}
\end{figure}

\begin{figure}[h]
\includegraphics[width=8cm]{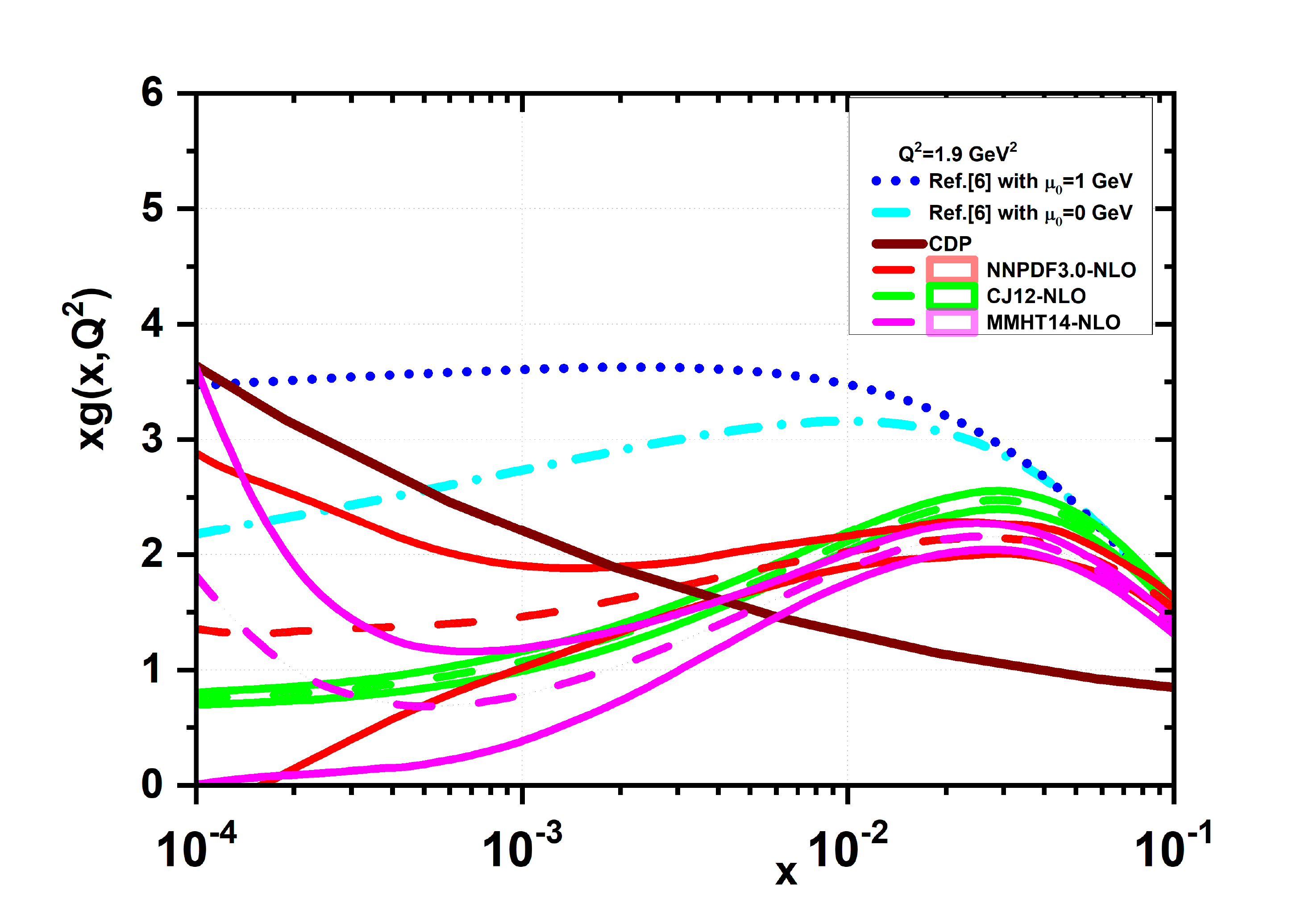}%
\caption{\label{Fig6}As Fig. 5, but for $xg (x,Q^2)$ instead of $\alpha_s
(Q^2) xg (x,Q^2)$. The CDP results are compared with the results from ref.[6] and the parametrization methods at the NLO approximation,  CJ12 \cite{CJ}, NNPDF3.0 \cite{NNPDF} and
 MMHT14 \cite{MMHT} as accompanied with total errors.
}
\end{figure}

For Figs. 5 and 6, the energy variable, $W$, of the
CDP is replaced by $x \simeq Q^2/W^2$. The increase of
$x$ for $Q^2 = 1.9~ {\rm GeV}^2$ in the range of 
$10^{-4} \le x \le 10^{-1}$ corresponds to an increase
of $\eta (W^2, Q^2 = 1.9~ {\rm GeV}^2)$ according to
$0.35 \le \eta \le 2.6$, with $W^2$ decreasing according
to $1.9 \times 10^4 {\rm GeV}^2 \ge W^2 \ge 19 {\rm GeV}^2$.
Taking into account the proportionality of the gluon distribution
to the longitudinal photoabsorption cross section (\ref{eq:3.31}) and consulting
the results in Fig. 1, we expect a (hadronlike) increase of the
gluon distribution by a factor of approximately about 3.5 with decreasing $x$
in the interval $10^{-1} \ge x \ge 10^{-4}$. This factor of 3.5 for $x=10^{-4}$  is seen in Figure 5.

Multiplication of the results in Fig. 5 by\break 
$\alpha_s (Q^2 = 1.9~
{\rm GeV}^2)^{-1} = 0.480^{-1} = 2.083$ yields the results for
$x g (x, Q^2) \equiv G (x, Q^2)$ that are shown in \break Fig. 6
\footnote{The value of $\alpha_s (Q^2 = 1.9~ {\rm GeV}^2) = 0.480$
is obtained for $\Lambda_{QCD} = 437 {\rm MeV}$ corresponding to
$\alpha_s (M^2_Z) = 0.118.$}.

The fairly small error bars of our results for the gluon distribution in
Fig. 5 are a reflection of the small errors of the original fit to the
experimental data of DIS.

The comparison of our CDP results for the gluon distribution in Fig. 6
and Fig. 10 below with published distributions reveals a significantly
different shape of the $x$ distribution at fixed $Q^2 = 1.9 {\rm GeV}^2$
and a large deviation in absolute normalization outside the total errors
of these distributions. Compare also the comment in the Conclusions
on the interpretation of the observed strong deviations.

In Fig. 6, we compare with the results obtained in ref. \cite{Pelicer}
upon introducing absorptive corrections to the measured proton structure
functions, and a modification of $\alpha_s (Q^2)$ by $\alpha_s (Q^2 + 
\mu^2_0)$. The gluon distributions from ref. \cite{Pelicer} in Fig. 6
are given by \cite{Pelicer}
\bqa
&&xg(x)=5.63x^{0.103}(1-x)^{10.261}~
\mathrm{for}~\mu_{0}=0~\mathrm{GeV},\nonumber \\
&&xg(x)=4.21x^{0.021}(1-x)^{9.427}~
\mathrm{for}~\mu_{0}=1~\mathrm{GeV}.
\label{eq:5.11}
\eqa

As seen in Fig. 6, the results from ref. \cite{Pelicer} are drastically
different from ours \footnote{Employing (2.8), we find that $F_{L}(x,Q^2)$ resulting by inserting the gluon distribution (5.11) differs from the CDP $F_{L}(x,Q^2)$ by the same factor as observed for the gluon distribution, see Fig.6.}. The results obtained by exploiting the smooth
low-$Q^2$ transition for $F_{L}(x,Q^2)$ of the CDP by substitution into the pQCD-improved
parton model do not support the  modifications of evolution suggested and applied to DIS in Ref.[6]. For further discussions on the consequences of our approach of
incorporating the CDP representation for $F_{L}(x,Q^2)$ into pQCD, see
sections VI and VII.

 For completeness, in Fig.7, we show the CDP gluon distribution for $Q^2$ between $Q^2=2~\textrm{GeV}^2$ and $Q^2=100~\textrm{GeV}^2$.

\begin{figure}[h]
\includegraphics[width=8cm]{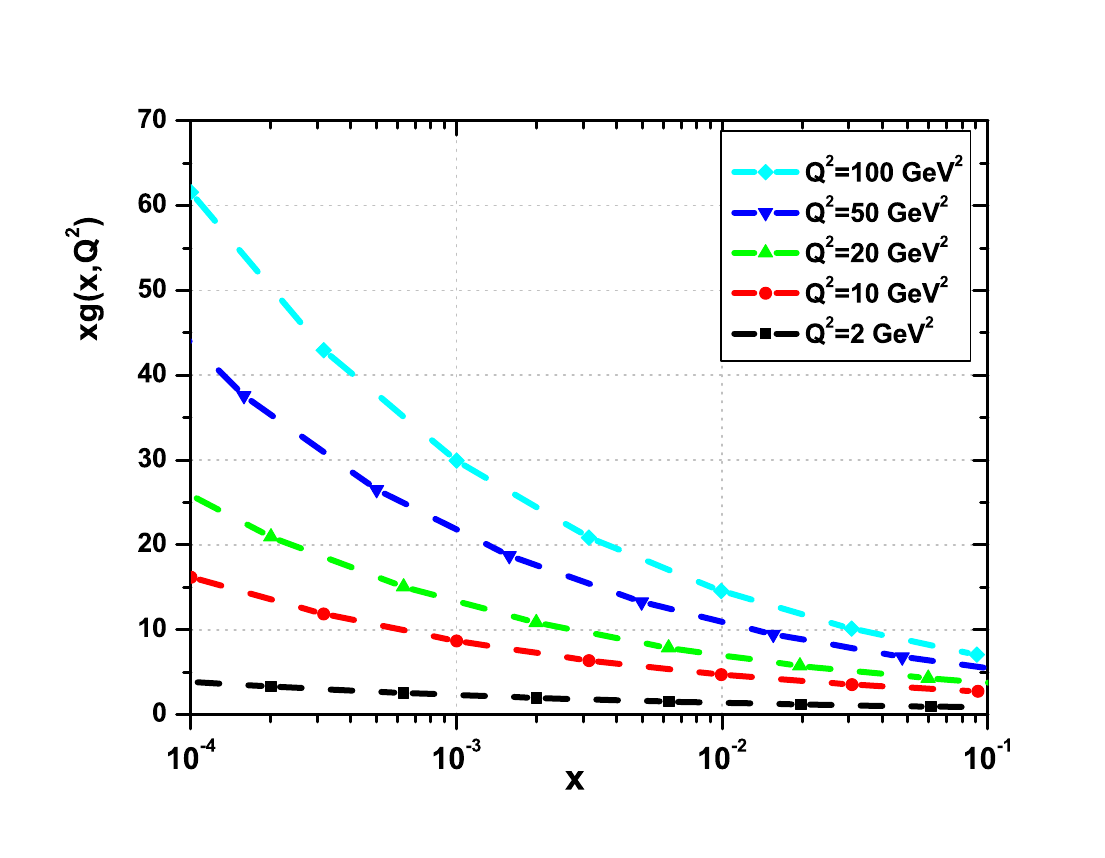}%
\caption{\label{Fig7}The CDP gluon distribution $xg (x,Q^2)$ in a wide range of $Q^2$ values, $2~\textrm{GeV}^2{\leq}Q^2{\leq}100~\textrm{GeV}^2$
}
\end{figure}

The analysis of the gluon distribution function that led to the results in Figs. 3
to 7 is based on the representation of the proton structure functions in the CDP. We expect
that other precise and theoretically well-founded representations (not manifestly including $q\bar{q}$ two-gluon couplings ) of the measured 
structure functions will lead to (at least approximately) identical results for
the gluon distribution. Explicitly this equivalence is tested by the example of
employing the fit to the proton structure function in the Froissart-bounded 
representation\footnote{We evaluate the gluon distribution from the pQCD approximation
(2.8) as well as (\ref{eq:6.14}) below, in distinction from the approach
in ref. \cite{Block} that is based on an exact determination of $G(x,Q^2)$ \cite{Durand}
upon converting (\ref{eq:6.10}) below into an inhomogeneous second order differential
equation for $G(x,Q^2)$.}
from ref. \cite{Block}, described in Section IV. 

Since the theoretical representation in this approach does not isolate the longitudinal
structure function, the gluon distribution must be deduced from $F_2 \left( \xi_L x,
Q^2 \right)$ with $\rho = \frac{4}{3}$ in (\ref{eq:5.10}).

\begin{figure}[h]
\includegraphics[width=8cm]{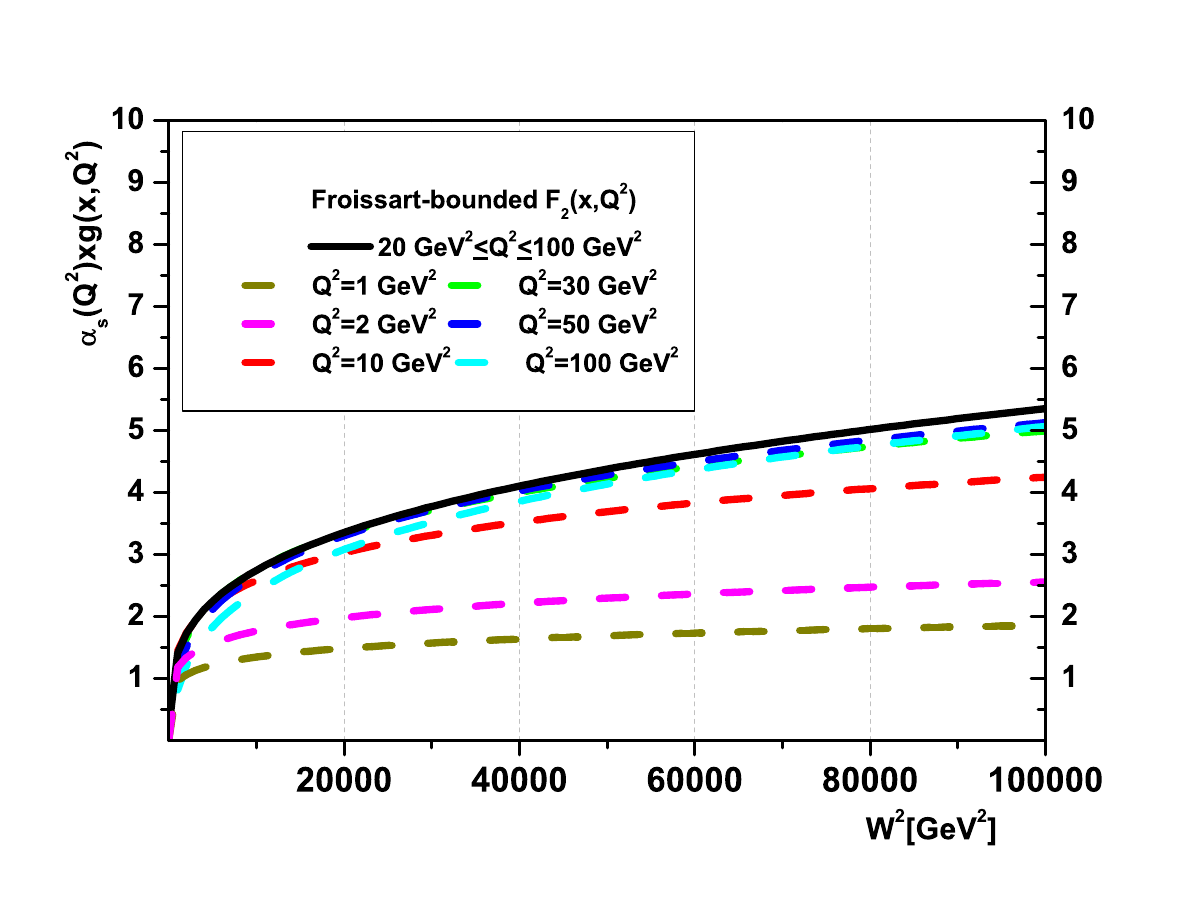}%
\caption{\label{Fig8}The gluon distribution $\alpha_s (Q^2) xg (x,Q^2)$ deduced from
the Froissart-bounded representation of the proton structure function $F_2 (x,Q^2)$
given in ref. \cite{Block}
}
\end{figure}

The results in Fig. 8  for large $Q^2$, asymptotically, agree with the CDP result. For
$Q^2 \lsim 10 ~{\rm GeV}^2$, there are acceptable deviations, seen upon comparing the 
results in Fig. 8 with the ones in Figs. 3 and 4.

\begin{figure}[h]
\includegraphics[width=8cm]{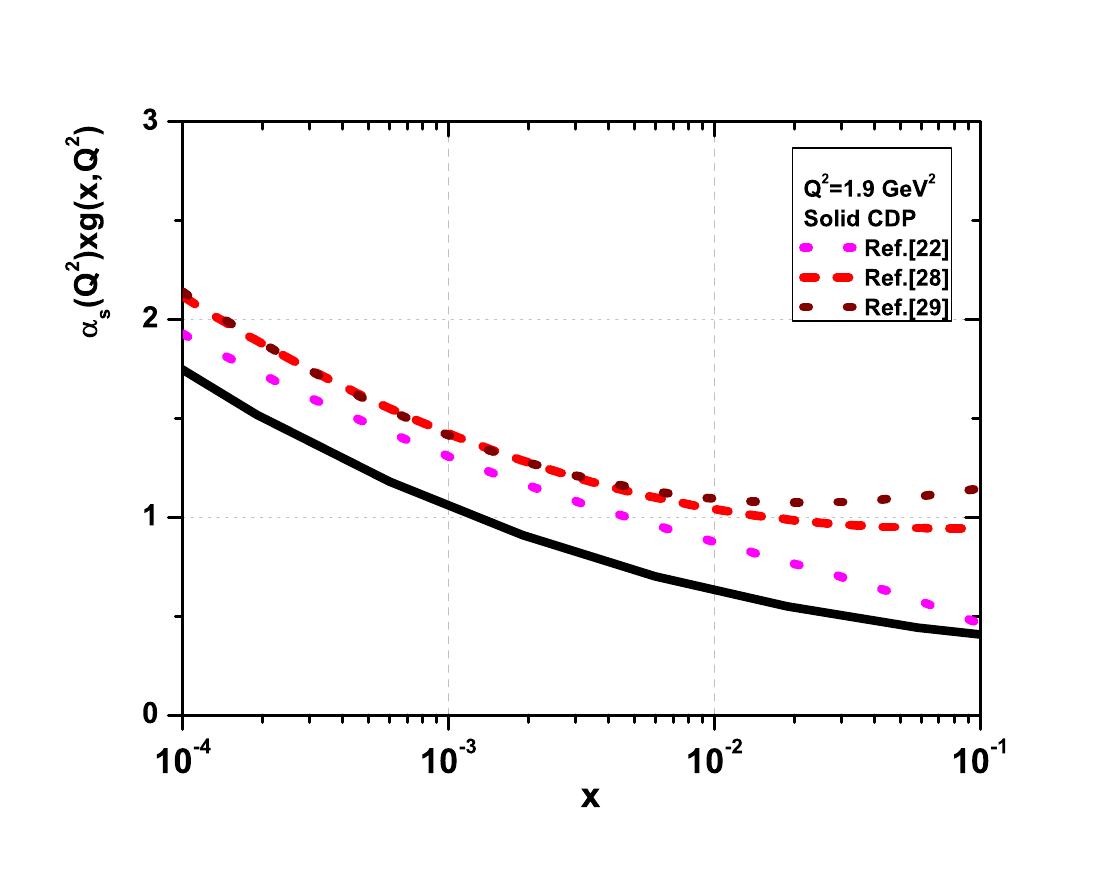}%
\caption{\label{Fig9}The gluon-distribution function $\alpha_s (Q^2) xg (x,Q^2)$
as a function of $x \cong Q^2/W^2$ at $Q^2 = 1.9 ~{\rm GeV}^2$ from the CDP, and
from Froissart-bounded representations in refs. \cite{Block}, \cite{Durand} and
\cite{Block-Durand}.
}
\end{figure}

\begin{figure}[h]
\includegraphics[width=8cm]{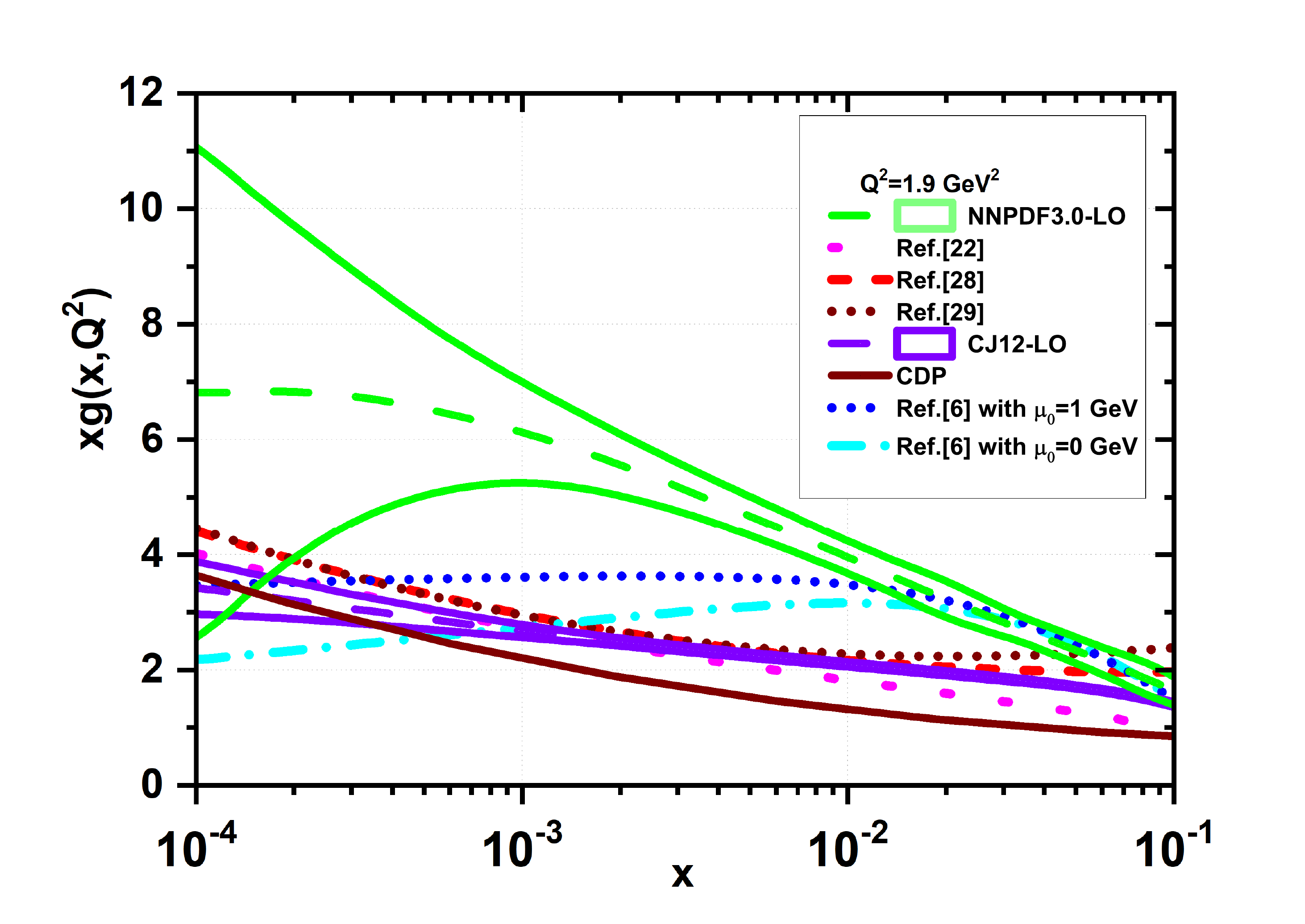}%
\caption{\label{Fig10}Same as Fig. 9 except for showing $xg(x,Q^2)$ instead of
$\alpha_s (Q^2) xg (x,Q^2)$. The CDP results and the results from the Froissart-bounded
representation are compared with the results from ref. \cite{Pelicer} and with the results CJ12 ref. \cite{CJ} and NNPDF3.0 ref. \cite{NNPDF} at the LO approximation as accompanied with total errors.
}
\end{figure}

In Figs. 9 and 10, respectively, we present the comparison of the gluon distribution from the 
Froissart-bounded representation of $F_2 (x,Q^2)$ with the results based on the CDP,		
and with the results from ref. \cite{Pelicer}
at $Q^2 = 1.9~ {\rm GeV}^2$ based on pQCD modified by power law corrections and absorptive
effects. 
In addition to the results extracted from the fit 
 to the
proton structure function described in Section IV., in Figs. 9 and 10, we have employed \cite{Block} two additional somewhat
different fits from refs. \cite{Durand, Block-Durand} without entering
into a detailed 
description of these fits. The CDP result in this figure (i.e., Fig.10) are compared with the CJ12 ref. \cite{CJ} and NNPDF3.0 ref. \cite{NNPDF} at the LO approximation. These parametrization methods are accompanied with total errors.

By comparison of the results on the gluon distribution deduced from the representation of
the body of DIS experimental data of the proton structure functions in the CDP, and the Froissart-bounded approach, at $Q^2 = 1.9~ {\rm GeV}^2$, we find
a consistent behavior as 
expected.

 The shape and the absolute magnitude of the CDP gluon distribution at $Q^2 = 1.9~{\rm GeV}^2$ is significantly different from all published results from several collaborations.

\section{VI. The Consistency of pQCD and CDP evolution at large $Q^2$}
\renewcommand{\theequation}{\arabic{section}.\arabic{equation}}
\setcounter{section}{6}\setcounter{equation}{0}

 In this Section, we examine the consistency
of the $Q^2$ evolution thus obtained for the gluon distribution from the CDP, with
the $Q^2$ dependence predicted from pQCD. To order $\alpha_s (Q^2)$, only 
process (2.5) contributes, and accordingly we can restrict
ourselves to investigating the consistency of the $Q^2$ dependence from the CDP with the
$Q^2$ dependence from the first DGLAP evolution equation \cite{Lipatov}.

The change of $F_2 (x, Q^2)$ with $Q^2$, its evolution at large $Q^2$,  is given by \cite{Lipatov}
\bqa
\frac{\partial}{\partial \ln Q^2} F_2 (x,Q^2) =\frac{\alpha_s (Q^2)}{2 \pi}
\int^1_x dz P_{qq} (z) F_2 \left( \frac{x}{z}, Q^2 \right) \label{eq:6.10}\nonumber \\
+\frac{R_{e^+e^-}}{3 \pi}\alpha_s (Q^2) \int^{1-x}_0 dz P_{qg} (z) G \left( \frac{x}{1-z}, Q^2 \right),~~~~~
\eqa
where
\be
P_{qq} (z) = \frac{4}{3} \left( \frac{1+x^2}{(1-x)_+} + \frac{3}{2} \delta (1-x) \right),
\label{eq:6.11}
\ee
and
\be
P_{qg} (z) = \frac{1}{2} \left( z^2 + (1-z)^2 \right).
\label{eq:6.12}
\ee
According to our basic assumption (2.7), saying that the photon exclusively interacts with $q \bar q$ pairs originating from a gluon, $g \to q \bar q$ see (2.6), we exclude the first term on the right-hand side in (6.1) to obtain  \cite{Prytz}
\bqa
\frac{\partial F_2 (x,Q^2)}{\partial \ln Q^2}= \frac{R_{e^+e^-}}{3 \pi} \int^{1-x}_0 dz P_{qg} (z) G \left( \frac{x}{1-z}, Q^2 \right).\nonumber\\
\label{eq:6.13}
\eqa
Exploiting the symmetry of $P_{qg} (z)$ in (6.4) around $z = 1/2$, the first
derivative of $G (x/(1-z), Q^2)$ in a Taylor expansion of $G (x/(1-z), Q^2)$ around
$z = \frac{1}{2}$ yields a vanishing contribution to the integral in (\ref{eq:6.13}),
and approximately \cite{Prytz}
\be
\frac{\partial F_2 (x,Q^2)}{\partial \ln Q^2} = \frac{R_{e^+e^-}}{9 \pi} \alpha_s (Q^2)
G \left(\frac{x}{\xi_2}, Q^2 \right).
\label{eq:6.14}
\ee
with $x/\xi_2 = 2 x$, or $\xi_2 = 1/2$. Replacing the gluon distribution in (\ref{eq:6.14})
by its proportionality (\ref{eq:2.4}) to the structure function $F_L $, (\ref{eq:6.14})
becomes
\be
\frac{\partial F_2(x,Q^2)}{\partial \ln Q^2} = F_L \left(\frac{\xi_L}{\xi_2} x, Q^2 \right) = 
\frac{1}{2 \rho + 1} F_2 \left(
\frac{\xi_L}{\xi_2} x, Q^2 \right),
\label{eq:6.15}
\ee
where the second equality is due to (3.18).

So far no specific ansatz for the proton structure function was used. Introducing the large-$Q^2$ power-law
of the CDP  (\ref{eq:3.21}) with (\ref{eq:3.10}),
\be
F_2 (x,Q^2) \sim (W^2)^{C_2} = \left( \frac{Q^2}{x} \right)^{C_2},
\label{eq:6.16}
\ee
from (\ref{eq:6.15}) we obtain the important constraint
\be
C_2 \left( \frac{Q^2}{x} \right)^{C_2} = \frac{1}{2 \rho +1} \left( 
\frac{Q^2}{x} \right)^{C_2} \left( \frac{\xi_2}{\xi_L} \right)^{C_2},
\label{eq:6.17}
\ee
or
\be
C_2 (2 \rho + 1) \left( \frac{\xi_L}{\xi_2} \right)^{C_2} = 1.
\label{eq:6.18}
\ee
Expansion of the exponential to first order in $C_2$ yields the 
convenient expression for $C_2$,
\be
C_2 \cong \frac{1}{2 \rho + 1} \frac{1}{\left( 1- \frac{1}{2 \rho + 1} \ln 
\frac{\xi_2}{\xi_L} \right) }.
\label{eq:6.19}
\ee
For $\xi_2/\xi_L = 0.5/0.4 = 1.25$, with $\rho = \frac{4}{3}$, one finds
$C_2 \cong 0.29$ for the exponent $C_2$ in (\ref{eq:6.18}).

The prediction of $C_2 \cong 0.29$ is based on the assumption that the
evolution in $Q^2$ is entirely due to $q \bar q$ pairs originating from gluons
$g \to q \bar q$, see
(6.4) and (6.5). The agreement of the prediction of $C_2{\cong}~ 0.29$ from (6.9)  with the experimentally determined value, given in (\ref{eq:3.20}), provides
empirical evidence for, or confirms the approximation of reducing the complete
evolution equation (6.1) to the reduced form  (6.5) employed to arrive at (6.9).
The incoming photon dominantly interacts with $q \bar q$ pairs originating from 
the $g \to q \bar q$ transition, see (2.6).

We conclude: the CDP with the power-law ansatz of $F_2 (x,Q^2) \sim (Q^2/x)^{C_2}$
for large $Q^2$ fulfills evolution according to (6.4) with $Q^2$ at large $Q^2$, and it implies the empirically successful
prediction of $C_2 \cong 0.29$.

\section{VII. Evolution at low $Q^2 \gsim 1.9 {\rm GeV}^2$.}
\renewcommand{\theequation}{\arabic{section}.\arabic{equation}}
\setcounter{section}{7}\setcounter{equation}{0}
 \bigskip
 
 In considering the validity of the pQCD evolution equation
 (6.5) for $F_2 (x,Q^2)$, we so far restricted ourselves to large values
 of $Q^2$, sufficiently large with respect to $\Lambda^2_{sat} (W^2)$, effectively,
 $Q^2 \gsim 10 {\rm GeV}^2$ to $20 {\rm GeV}^2$. Employing the large-$Q^2$-representation
 for $F_2 (x, Q^2) \sim \left( \frac{Q^2}{x} \right)^{C_2}$, the validity of the evolution
 equation (6.6) implied the restriction (\ref{eq:6.18}) that predicts the exponent
 $C_2$ in $\Lambda^2_{sat} (W^2) \sim (W^2)^{C_2}$ in terms of the transverse-$q \bar q$-size parameter
 $\rho$, where $\rho$ has the preferred value [24] of $\rho = 4/3$.
 
 In the present Section, we lift the restriction on the magnitude of $Q^2$ by allowing for
 values of $Q^2$ as low as $Q^2 \gsim 1.9 {\rm GeV}^2$, the value frequently adopted [6] when
 applying pQCD to electron-proton DIS.
 
 The CDP proton structure functions entering (\ref{eq:6.15}) according to (\ref{eq:3.1}), 
 (\ref{eq:3.2}) and (\ref{eq:3.12}), without restriction to large $Q^2$ are given by
 
 \bqa
 & F_2 (x,Q^2) \equiv F_2 \left( \eta \left( W^2 = \frac{Q^2}{x}, Q^2 \right), \mu \left( 
 W^2 = \frac{Q^2}{x} \right) \right) \nonumber\\ 
  & = \frac{Q^2}{4 \pi^2 \alpha} \frac{\sigma_{\gamma p} (W^2)}{\ln \frac{\rho}{\mu (W^2)}}
 \left( I_T^{(1)} \left( \frac{\eta}{\rho}, \frac{\mu}{\rho} \right) G_T (u) +
 I^{(1)}_L (\eta, \mu) G_L (u) \right),\nonumber\\
& \label{eq:7.1}
\eqa
 and
 \bqa
 F_L (x,Q^2) &\equiv F_L \left( \eta \left( W^2 = \frac{Q^2}{x}, Q^2 \right), \mu \left( 
 W^2 = \frac{Q^2}{x} \right) \right) \nonumber \\
 & = \frac{Q^2}{4 \pi^2 \alpha} \frac{\sigma_{\gamma p} (W^2)}{\ln \frac{\rho}{\mu (W^2)}}
 I_L^{(1)}  G_L (u). 
 \label{eq:7.2}
 \eqa
The fit to the experimental data of
$\sigma_{\gamma p} (W^2)$ is given by (\ref{eq:3.7}). For the explicit expressions
for $I_T^{(1)}$ and $I_L^{(1)}$ in (\ref{eq:7.1}) and (\ref{eq:7.2}), we refer to Appendix A.
In the present context we can restrict ourselves to the range of $\eta \ll \xi = 130$, and
accordingly $G_{L,T} (u) = 1$ in (\ref{eq:7.1}) and (\ref{eq:7.2}), see (\ref{eq:3.4}) and
(\ref{eq:3.5}).

Adopting (\ref{eq:3.10}),
\be
\Lambda^2_{sat} (W^2) = C_1 \left( \frac{W^2}{1 {\rm GeV}^2} \right)^{C_2},~~~W^2 = \frac{Q^2}{x},
\label{eq:7.3}
\ee
the low-$x$ scaling variable $\eta (W^2,Q^2)$ from (\ref{eq:3.8}),  expressed in terms of the parton variables $x$ and $Q^2$, 
becomes
\be
\eta \equiv \eta (W^2=\frac{Q^2}{x},Q^2) = \frac{x^{C_2}}{C_1} \frac{(Q^2 + m^2_0)}{(Q^2)^{C_2}},
\label{eq:7.4}
\ee
and $\mu (W^2)$ from (\ref{eq:3.9}) is given by
\be
\mu \equiv \mu (W^2=\frac{Q^2}{x}) = \frac{x^{C_2}}{C_1} \frac{m^2_0}{(Q^2)^{C_2}}.
\label{eq:7.5}
\ee
The numerical value of the exponent $C_2$, for a given value of $\rho$ is fixed by the large-$Q^2$ constraint (\ref{eq:6.18}).

Turning to the examination of the validity of the evolution equation for $F_2 (x, Q^2)$
in the form of the first equality in (\ref{eq:6.15}) relating the logarithmic derivative of $F_{2}(x,Q^2)$ to the longitudinal structure function $F_{L}(x,Q^2)$, we introduce the ratio
\bqa
{\rm Ratio} && \equiv \frac{\frac{\partial F_2 (x,Q^2)}{\partial \ln Q^2}}
{F_L \left( \frac{\xi_L}{\xi_2} x, Q^2 \right)} \label{eq:7.6} \\
&& = 1 + \Delta
\left( \eta \left( W^2 = \frac{Q^2}{x}, Q^2\right), \mu
\left( W^2 = \frac{Q^2}{x} \right) \right),  \nonumber
\eqa
where, according to the validity of (\ref{eq:6.15}) under constraint
(\ref{eq:6.18}), we have $\Delta (\eta, \mu) \cong 0$ for sufficiently
large $Q^2 \gsim 10 {\rm GeV}^2$. Deviations from the validity of the
evolution equation (6.5) are accordingly parametrized by $\Delta (\eta, \mu)$
according to (\ref{eq:7.6}).

Note that  $\Delta (\eta, \mu){ \neq}0$ in (7.6) may be due to either a consequence of a violation of the evolution equation (6.5), under validity of the relation between the gluon distribution and the longitudinal structure function (2.8), or else to a violation of (2.8), not excluding both cases simultanusely. In any case, $\Delta (\eta, \mu){ \neq}0$ implies violation of the underlying parton model.  At low $\eta$, the interaction is dominantly due to the interaction of low-lying $q \bar q$ vector states with the gluon field in the proton, or, equivalently (see(2.6)) to the interaction of the photon with low-lying $q \bar q$ vector states and not with freely moving quarks and gluons. Nevertheless, one may formally define a gluon distribution function according to (2.9) and (3.32). 

From (\ref{eq:7.1}), the derivative of $F_2 (x, Q^2)$ is obtained as
\bqa
\frac{\partial F_2 (x,Q^2)}{\partial \ln Q^2} &=& \frac{Q^2}{4 \pi^2 \alpha}
\frac{\sigma_{\gamma p} (W^2)}{\ln \frac{\rho}{\mu (W^2)}} \left[
\left( I_T^{(1)} \left( \frac{\eta}{\rho}, \frac{\mu}{\rho} \right) +
I_L^{(1)} (\eta, \mu) \right)  \right.
\nonumber \\
&\times & \left( 1 + \frac{\ln \frac{\rho}{\mu (W^2)}}{\sigma_{\gamma p} (W^2)}
\frac{\partial}{\partial \ln W^2} \frac{\sigma_{\gamma p} (W^2)}{\ln
\frac{\rho}{\mu (W^2)}} \right) \nonumber \\
& + & \left. \frac{\partial}{\partial \ln Q^2} \left( I^{(1)}_T \left( \frac{\eta}{\rho},
\frac{\mu}{\rho} \right) + I_L^{(1)} (\eta, \mu) \right) \right],
\label{eq:7.7}
\eqa

and the denominator of (\ref{eq:7.6}) is given by
\be
F_L \left( \frac{\xi_L}{\xi_2} x, Q^2 \right) = \frac{Q^2}{4 \pi^2 \alpha}
\frac{\sigma_{\gamma p} \left( \frac{\xi_2}{\xi_L} W^2 \right)}{\ln
\frac{\rho}{\mu \left(\frac{\xi_2}{\xi_L} W^2 \right)}} I_L^{(1)}
\left( \frac{\xi_L}{\xi_2} \eta, \frac{\xi_L}{\xi_2} \mu \right).
\label{eq:7.8}
\ee
In connection with the derivatives appearing in (\ref{eq:7.7}), we mention
the equality $\frac{\partial}{\partial \ln W^2} = \frac{\partial}{\partial \ln Q^2}$,
and moreover, we note the derivatives
\bqa
Q^2 \frac{\partial \eta}{\partial Q^2} &= (1-C_2) \eta - \mu, \nonumber \\
Q^2 \frac{\partial \mu}{\partial Q^2} & = - C_2 \mu.
\label{eq:7.9}
\eqa

Since  $\sigma_{\gamma p} (W^2)$ only weakly depends on $W^2$, one may 
approximate ``Ratio'' defined in (\ref{eq:7.6}) by ignoring the derivative
of $\sigma_{\gamma p} (W^2)$ in (\ref{eq:7.7}). Explicitly, Ratio from
(\ref{eq:7.6}) becomes\footnote{In (\ref{eq:7.10}), the shift $\xi_2/\xi_L W^2$
in $\sigma_{\gamma p} \left( \frac{\xi_2}{\xi_L} W^2\right) / \ln 
\frac{\rho}{\mu \left( \frac{\xi_2}{\xi_L} W^2 \right)}$ from (\ref{eq:7.8}) is
ignored.}
\bqa
& Ratio = \frac{1}{I^{(1)}_L \left( \frac{\xi_L}{\xi_2}\eta, 
\frac{\xi_L}{\xi_2} \mu \right)} \label{eq:7.10}  \\
& \times  \left[ I_T^{(1)} \left( \frac{\eta}{\rho}, \frac{\mu}{\rho} \right)
+ I^{(1)}_L (\eta, \mu) \right. \nonumber \\
& \left. + \frac{\partial}{\partial \ln Q^2}
\left(I_T^{(1)} \left(\frac{\eta}{\rho}, \frac{\mu}{\rho} \right) +
I_L^{(1)} (\eta, \mu) \right) \right]\nonumber \\
&= \frac{1}{I^{(1)}_L \left( \frac{\xi_L}{\xi_2}\eta, 
\frac{\xi_L}{\xi_2} \mu \right)} \times  \left[ I_T^{(1)} \left( \frac{\eta}{\rho}, \frac{\mu}{\rho} \right)
+ I^{(1)}_L (\eta, \mu) \right. \nonumber \\
& \left. + [(1-C_{2})\eta-\mu]\frac{\partial}{\partial  \eta}
\left(I_T^{(1)} \left(\frac{\eta}{\rho}, \frac{\mu}{\rho} \right) +
I_L^{(1)} (\eta, \mu) \right) \right].\nonumber
\eqa
Moreover, since $\mu (W^2) = m^2_0/\Lambda^2_{sat} (W^2)$, with $m^2_0 = 0.15
{\rm GeV}^2$, for sufficiently large $W^2, \mu (W^2)$ is small compared to unity,
$\mu (W^2) \ll 1$, and the approximation of $I_L^{(1)}$ and $I_T^{(1)}$ in terms of 
$I_0 (\eta)$ \cite{Kuroda}   given by (A3) to (A5) in Appendix A may be inserted when numerically
evaluating Ratio in (\ref{eq:7.6}) and (\ref{eq:7.10}).

Making use of
\be
\frac{d}{d\eta} I_0(\eta) = \frac{-2}{1+4 \eta} \left(I_0 (\eta) - \frac{1}{2 \eta} \right)
\label{eq:7.11}
\ee
together with (\ref{eq:7.9}), Ratio in (\ref{eq:7.10}) becomes a function that depends
on $\eta$, $\mu$ and $I_0(\eta)$. The different dependence of $I_0 (\eta)$ on $\eta$ at large
$\eta \gg 1$, where
\be
I_0 (\eta) \cong \frac{1}{2 \eta} \left(1- \frac{1}{6 \eta} \right) + 0 
\left( \frac{1}{\eta^3}\right),
\label{eq:7.12}
\ee
and at small $\eta$ of $\mu \le \eta \ll 1$, where
\be
I_0 (\eta) \cong \ln \frac{1}{\eta} \left( 1 - 2 \eta \left(1 - \frac{1}{\ln \frac{1}{\eta}} \right)
\right) + ... ,
\label{eq:7.13}
\ee
determines the different behavior of the numerical results for Ratio to be presented next.

\begin{figure}[h]
\includegraphics[width=8cm]{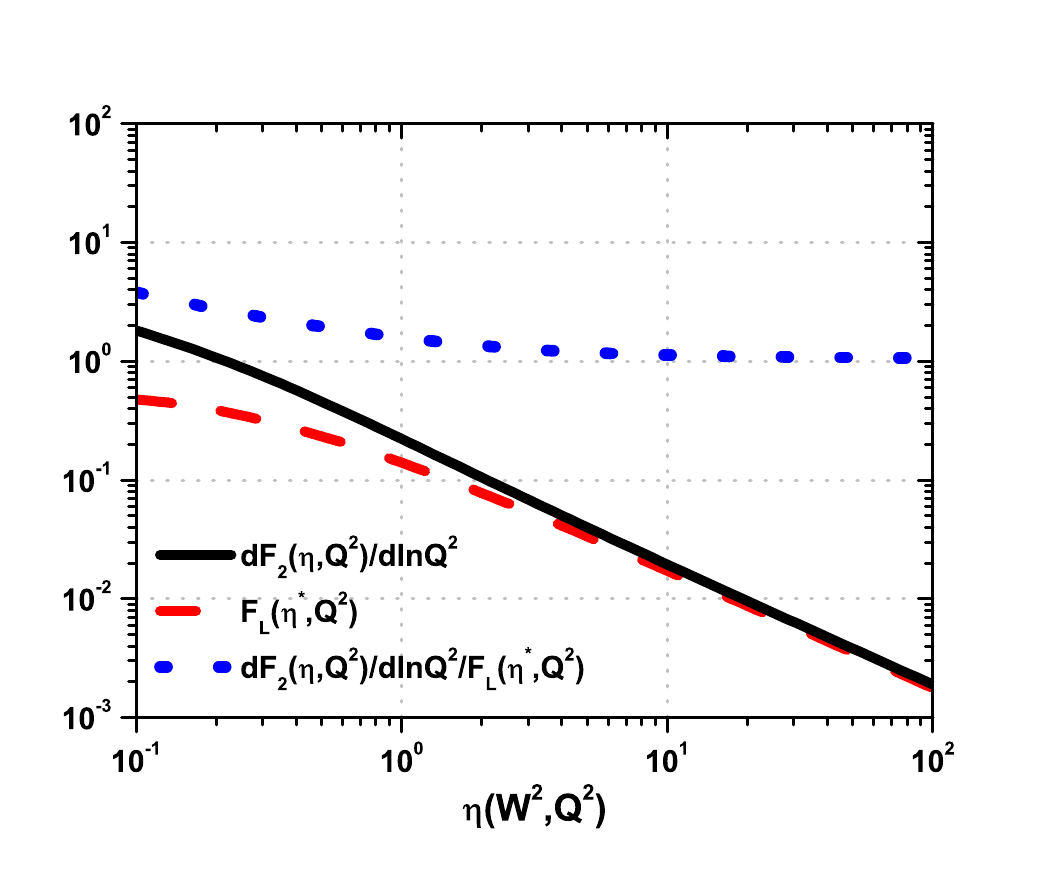}%
\caption{\label{Fig10}The derivative of the structure function $F_2 (\eta (W^2 = Q^2/
x),Q^2)$ and the longitudinal structure function $F_L (\eta^{*}=(\frac{\xi_{L}}{\xi_{2}})^{C_{2}}\eta, Q^2)$
together with the ratio of both of them. The results are based on evaluating (\ref{eq:7.7})
and (\ref{eq:7.8}).}
\end{figure}

For the numerical evaluation of Ratio from (\ref{eq:7.6}) with (\ref{eq:7.7}) and
(\ref{eq:7.8}) inserted, and of Ratio explicitly given in (\ref{eq:7.10}), we use
the parameters\footnote{We add the comment that only the value of $C_1 = 0.31
{\rm GeV}^2$ is an entirely free fit parameter, whereas $\rho = \frac{4}{3}$ is
a consequence of the enhanced transverse size of $q \bar q$ states originating
from transversely relative to longitudinally polarized photons, and the
exponent $C_2 = 0.29$ is accordingly fixed by (\ref{eq:6.18}). The value of
$m^2_0 = 0.15 {\rm GeV}^2 < m^2_\rho$ is the starting point of the effective
$q \bar q$ continuum, including the $\rho^0, \omega, \phi$ peaks. We add the remark that $\sqrt{0.31}\rm{GeV}{\cong}0.57\rm{GeV}$ may be associated with the level spacing between the vector   mesons $\rho^0$ and $\acute{\rho}^{0}$.}
\be
\rho = \frac{4}{3},~C_2 = 0.29,~C_1 = 0.31 ~{\rm GeV}^2,~m^2_0 = 0.15 {\rm GeV}^2.
\label{eq:7.14}
\ee
We also note $\xi_L = 0.4$ (see (\ref{eq:2.4})) and $\xi_2 = 0.5$ (see (\ref{eq:6.14})).\\

In Fig. 11, we show the derivative of the structure function as well as the longitudinal
structure function according to (\ref{eq:7.7}) and (\ref{eq:7.8}) respectively, and their
ratio defined by (\ref{eq:7.6}). For $W^2$, the value of $W^2 = 10^5~{\rm GeV}^2$ was
used \footnote {As a consequence of the dominant dependence on $\eta(W^2,Q^2)$, the absolute value of $W^2$ in only relevant with respect to $\sigma_{\gamma p}(W^2)$ and $\mu(W^2)$ in (7.7) and (7.8). }.

\begin{figure}[h]
\includegraphics[width=8cm]{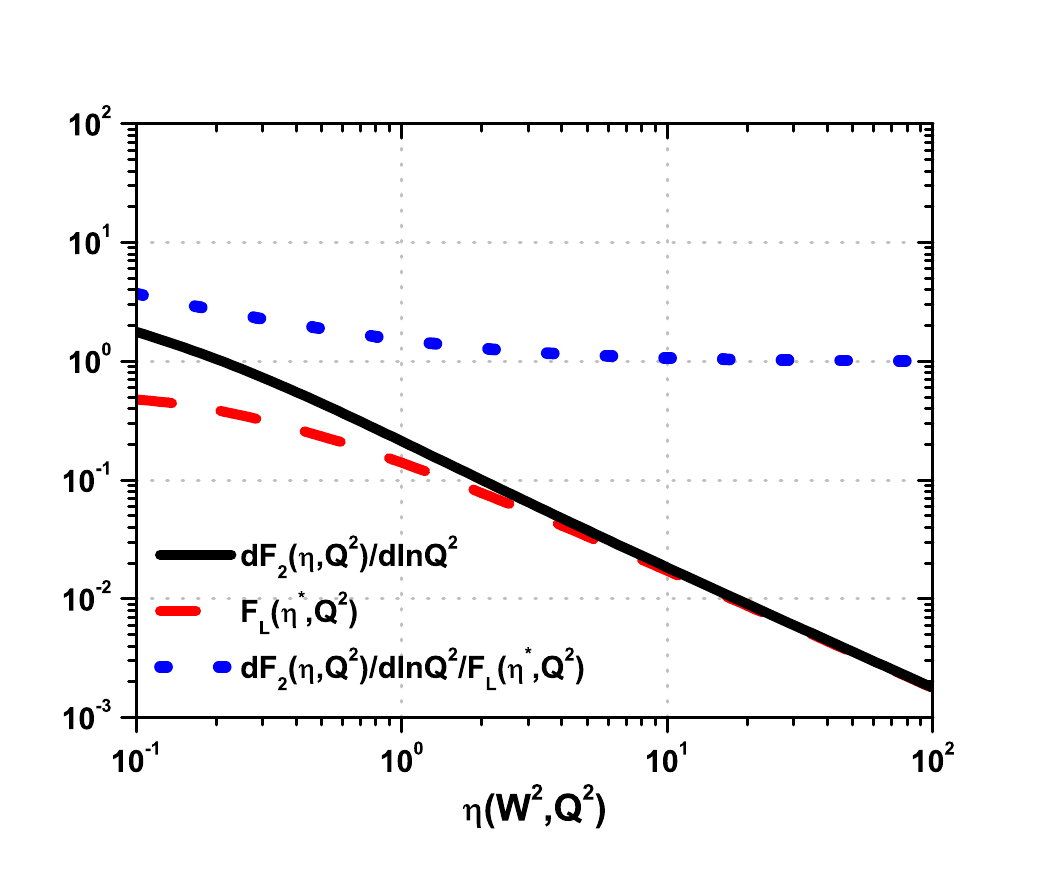}%
\caption{\label{Fig11}As Fig. 11, however using the approximation
$\sigma_{\gamma p} (W^2) \cong {\rm const}$, compare to (\ref{eq:7.10}).
}
\end{figure}
\begin{figure}[h]
\includegraphics[width=8cm]{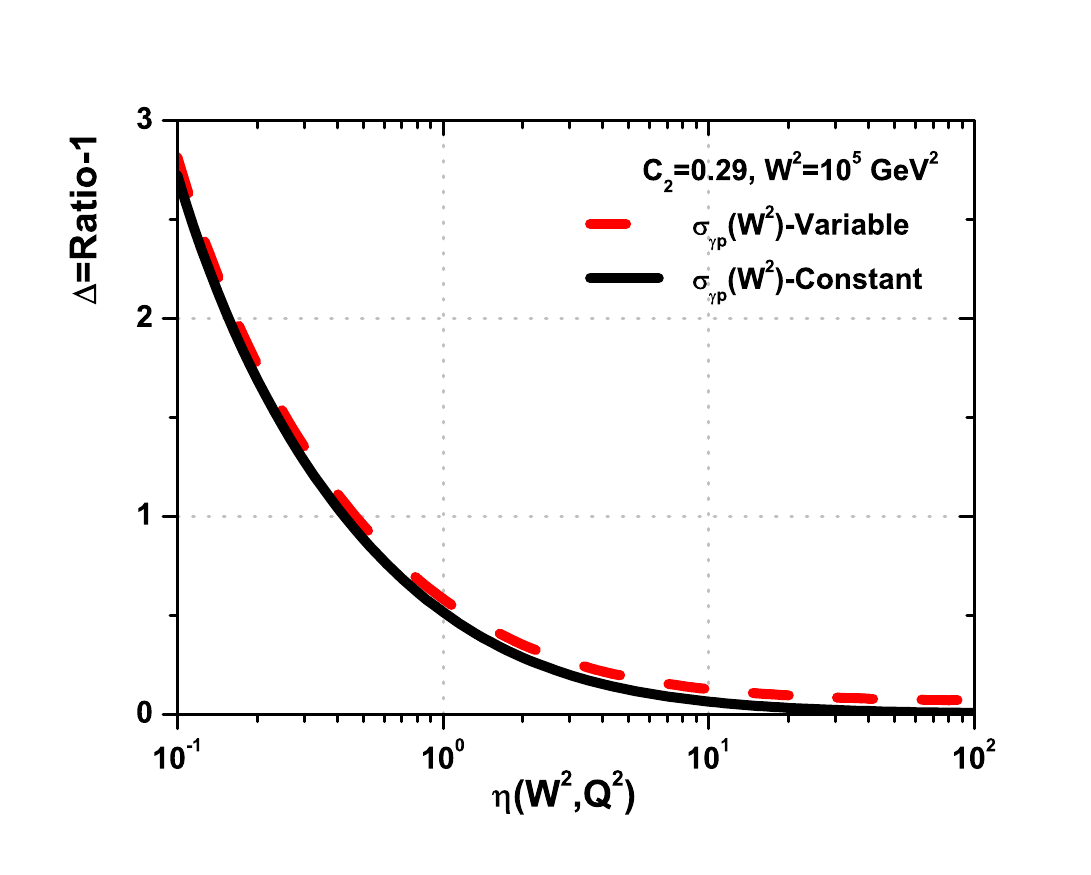}%
\caption{\label{Fig12}The deviation from validity of the evolution equation
obtained according to Figs. 11 and 12, respectively.
}
\end{figure}

Since $\sigma_{\gamma p} (W^2)$ depends weakly on $W^2$, we expect the numerical
results of Fig. 11 not being drastically affected, if in (\ref{eq:7.7}) the term
proportional to the derivative of $\sigma_{\gamma p} (W^2)$ is put to zero. The
corresponding value of Ratio is explicitly given by (\ref{eq:7.10}). 
Comparison of the results in Fig. 12 with the ones
in Fig. 11 indeed confirms that Ratio can reliably be evaluated according to
the simple expression (\ref{eq:7.10}) upon insertion of $I_T^{(1)}$ and $I_L^{(1)}$
from (A.3) to (A.5).

In Fig. 13, we show the deviation
 $\Delta \left( \eta \left( W^2 = \frac{Q^2}{x} \right),
\mu (W^2) \right)$
of Ratio in (\ref{eq:7.6}) from unity on a linear scale. 
Remember, a value of unity means strict validity of the first
DGLAP evolution equation (6.1) in the approximation (6.4) and
(6.5). A deviation smaller than $\Delta \left( \eta \left( W^2 = \frac{Q^2}{x} \right),~
\mu (W^2) \right) \cong 0.1$ to $0.2$ according to Fig. 13 requires $\eta (W^2,Q^2)
\gsim 5$. The weaker increase, and finally the decrease of $F_L (\eta (x, Q^2), Q^2)$ relative to the derivative of $F_2 (\eta (x, Q^2), Q^2)$
with decreasing $\eta (x, Q^2)$, or decreasing $x$ at fixed $Q^2$, see (7.4),  implies the substantial violation of the evolution
equation reaching $\Delta (\eta (W^2, Q^2), \mu (W^2)) \cong 2.7$ at
$\eta (W^2,Q^2) \cong 0.1$.

\begin{figure}[h]
\includegraphics[width=8cm]{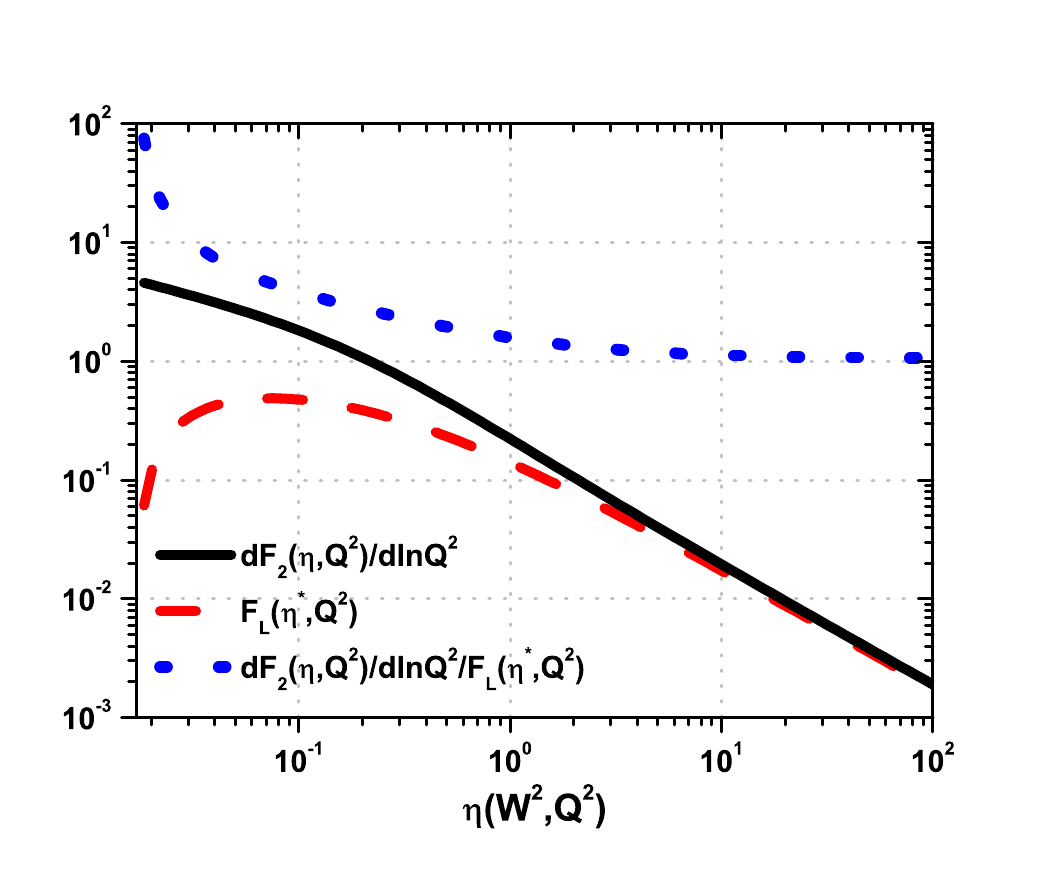}%
\caption{\label{Fig13}The extension of Fig. 11 to the $Q^2 \to 0$ limit of
$\eta (W^2,Q^2) \to \mu (W^2)$ for $W^2 = 10^5 {\rm GeV}^2$
}
\end{figure}

Fig. 14 explicitly shows that the increase of Ratio relative to unity with decreasing
$\eta$ is strongly correlated to the necessary vanishing of the longitudinal structure function,
$F_L (\eta (W^2,Q^2), \mu (W^2))$, in the denominator of Ratio in the limit of $\eta (W^2,Q^2)
\to \mu (W^2)$ for $Q^2 \to 0$.

\begin{figure}[h]
\includegraphics[width=8cm]{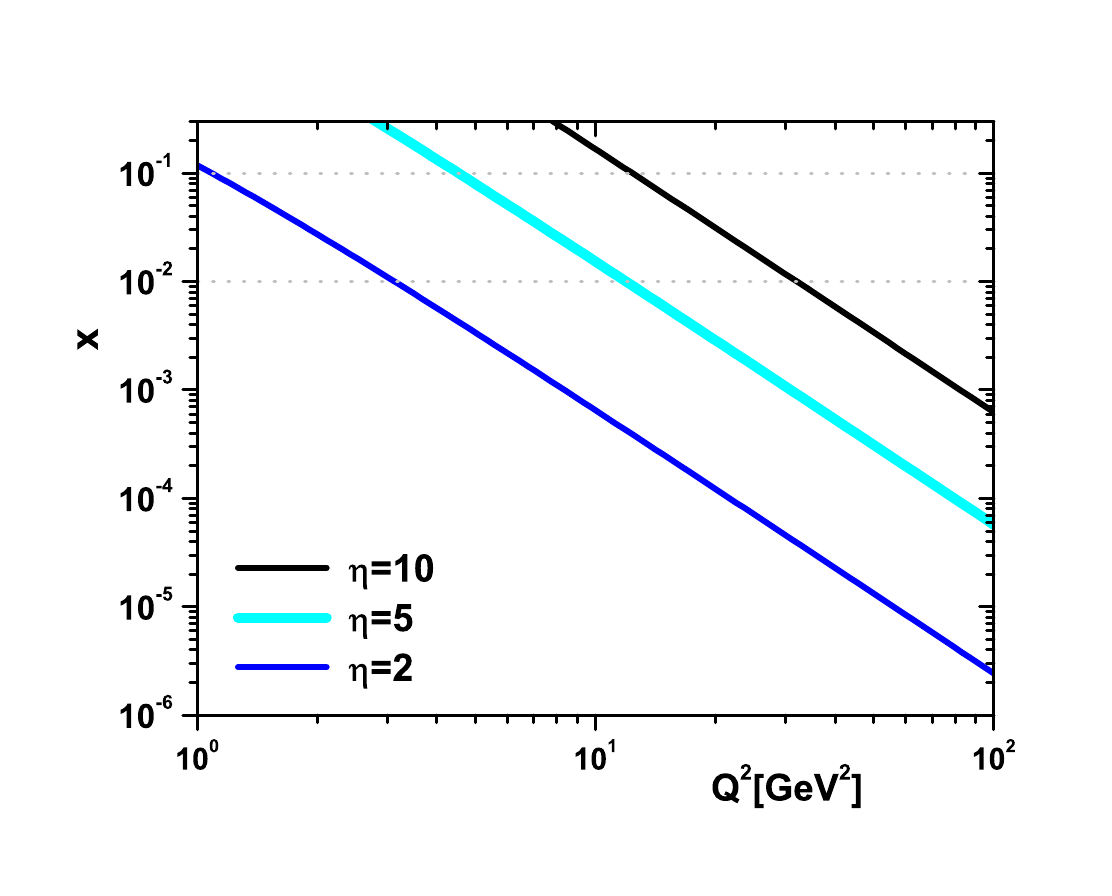}%
\caption{\label{Fig14}The $(Q^2,x)$ plane, showing the regions of $\eta (W^2,Q^2) > 5$
(approximate validity of conventional evolution) with $\Delta (\eta, \mu) \cong 0$,
and of $\eta (W^2,Q^2) < 5$, where $\Delta (\eta, \mu) \not= 0$ occurs.
}
\end{figure}

The value of $\eta \cong 5$ quantifies the region in the $(Q^2,x)$ plane, see Fig. 15,
below which the evolution equation becomes violated. It is amusing to note the coincidence
of this value with the value of $\eta \cong 5$, where the behavior
of $\sigma_{\gamma^*p} \sim 1/\eta$ of the CDP in Fig. 1 turns from the region of color
transparency to the region of hadronlike behavior. Above $\eta \cong 5$, we have
pQCD-evolution or CDP-color-transparency, while below $\eta \cong 5$, we have a 
violation of conventional evolution that is equivalent to hadronlike saturation.

\begin{figure}[h]
\includegraphics[width=8cm]{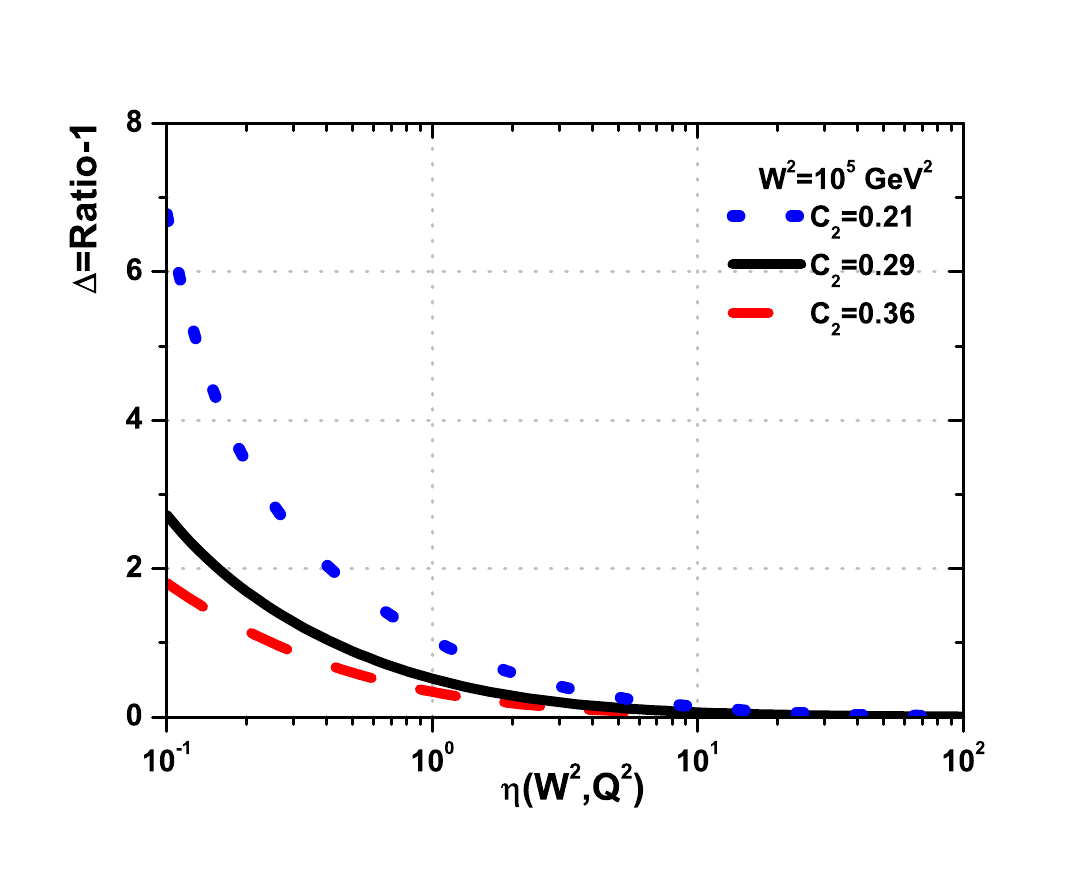}%
\caption{\label{Fig15}The effect of the variation of $\rho$.
}
\end{figure}

In terms of $x$ and $Q^2$, according to Fig.15, at fixed value $Q^2$ with decreasing $x$,  or at fixed value of $x$ with decreasing $Q^2$, the region of hadronlike behavior is reached that in connected with a violation of standard evolution.

In Fig. 16, we address the question of the dependence of the above conclusion on the 
value of $\rho$ that specifies the asymptotic behavior of the longitudinal-to-transverse
ratio in photoabsorption via $R = 1/2 \rho$. As long as precision experimental data on $R$,
and accordingly  on $\rho$, are lacking, it is interesting to vary $\rho$ around its
preferred value of $\rho = 4/3$, implying different values of the exponent $C_2$
according to (\ref{eq:6.18}). From Fig. 16, we infer that a change of $\rho$ does 
qualitatively not
significantly change our general conclusion obtained for $\Delta (\eta (W^2 = Q^2/x), Q^2)$.

In Fig.17, we show Ratio according to (7.10) as a function of $Q^2$ at fixed $x$, upon replacing $\eta$ and $\mu$ in terms of $x$ and $Q^2$ according to (7.4) and (7.5). The figure shows for decreasing fixed values of $x$ the increasingly stronger deviation of Ratio  from Ratio=1  with decreasing $Q^2$.

\begin{figure}[h]
\includegraphics[width=8cm]{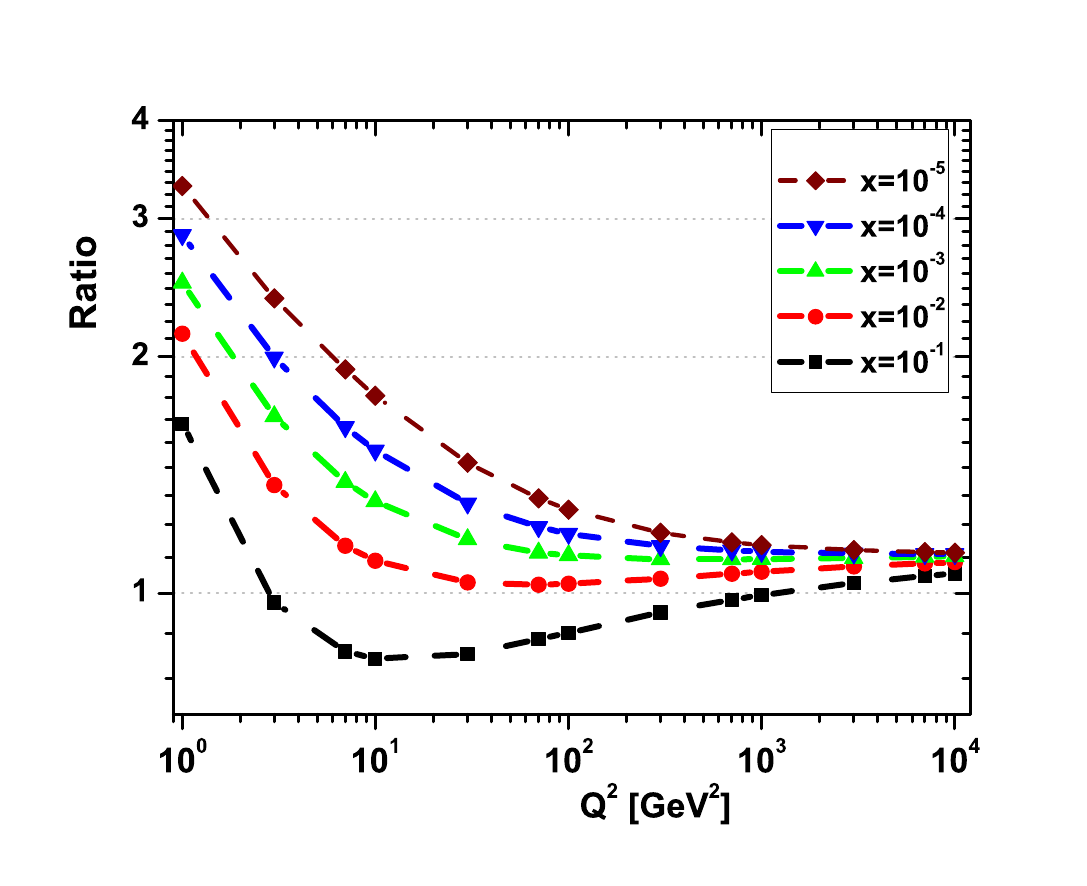}%
\caption{\label{Fig7}The Ratio  (7.6) as a function of $Q^2$  in a wide range of x, $x=10^{-5...-1}$.
}
\end{figure}

In Figs. 11 to 14, we showed Ratio from (\ref{eq:7.6}) as a function of $\eta (W^2,Q^2)$. In connection with the extraction of the gluon distribution from DIS experimental data, one
frequently uses a fixed value of $Q^2$, chosen as low as $Q^2 = 1.9~ {\rm GeV}^2 \cong
2~ {\rm GeV}^2$, as starting scale for the evolution of the gluon distribution with increasing
$Q^2$ at fixed $x$.

\begin{table}[h]
\centering \caption{Here $\Delta=\mathrm{Ratio}-1$ and
$C_{2}=0.29$ .}\label{table:table2}
\begin{minipage}{\linewidth}
\renewcommand{\thefootnote}{\thempfootnote}
\centering
\begin{tabular}{|l|c||c|c|c|c|} \hline\noalign{\smallskip}  $Q^{2}[\mathrm{GeV}^{2}]$ & $ x
$ &
$W^{2}[\mathrm{GeV}^{2}]$ & $\eta$ & $\Delta+1$ & $\frac{1}{\Delta+1}$  \\
\hline\noalign{\smallskip}
2 & $10^{-2}$ & 2.$10^2$ & 1.492 & 1.544 & 0.648 \\
2 & $10^{-3}$ & 2.$10^3$ & 0.765 & 1.894 & 0.528 \\
2 & $10^{-4}$ & 2.$10^4$ & 0.392 & 2.252 & 0.444 \\
2 & $10^{-5}$ & 2.$10^5$ & 0.201 & 2.662 & 0.376 \\
\hline\noalign{\smallskip}
20 & $10^{-2}$ & 2.$10^3$ & 7.172 & 1.041 & 0.961 \\
20 & $10^{-3}$ & 2.$10^4$ & 3.678 & 1.204 & 0.831 \\
20 & $10^{-4}$ & 2.$10^5$ & 1.886 & 1.362 & 0.734 \\
20 & $10^{-5}$ & 2.$10^6$& 0.967 & 1.563 & 0.640 \\
\hline\noalign{\smallskip}
\hline\noalign{\smallskip}
100 & $10^{-2}$ & 1.$10^4$ & 22.351 & 1.026 & 0.975 \\
100 & $10^{-3}$ & 1.$10^5$ & 11.463 & 1.115 & 0.897 \\
100 & $10^{-4}$ & 1.$10^6$& 5.879 & 1.189 & 0.841 \\
100 & $10^{-5}$ & 1.$10^7$ & 3.015 & 1.277 & 0.783 \\
\hline\noalign{\smallskip}
\end{tabular}
\end{minipage}
\end{table}

In Table III, we show the results for $(1 + \Delta)^{-1}$ according to
(\ref{eq:7.6}) with (\ref{eq:7.7}) and (\ref{eq:7.8}) for fixed $Q^2 = 2~ {\rm GeV}^2$ and
the typical range of $10^{-2} \le x \le 10^{-5}$, corresponding to an interval of $\eta$
approximately given by $1.5 \ge \eta \ge 0.2$. According to Table III, we observe a dramatic
correction factor to standard evolution of magnitude $0.6 \ge (1 + \Delta)^{-1} \ge 0.4$.
This correction to evolution is due to a strong violation of the impulse approximation
of the  pQCD improved parton model at low $Q^2$. The associated transition to hadronlike $(q \bar q) p$ interactions
is outside the range of validity of the standard evolution equations. It comes without
surprise that global DIS data fits based on imposing evolution from a low-$Q^2$-starting
input scale ``do not describe the deep inelastic scattering data in the low-$x$, low-$Q^2$
region very well'', see ref. \cite{Pelicer}. We add a comment on the interpretation of the huge
discrepancy between the CDP gluon distributions in Figs.6 and 10 and the published results
from several collaborations. The CDP gluon distribution is deduced from color gauge-invariant
CDP structure functions explicitly incorporating the $Q^2{\rightarrow}0$ limit.
The standard determinations contain a relatively ad hoc continuation to low values of $Q^2$. 

In Table III, in addition to the choice of $Q^2 = 2~ {\rm GeV}^2$, we also give the results
for the correction factor $(1 + \Delta)^{-1}$ at $Q^2 = 20 {\rm GeV}^2$. For $x = 10^{-2}$,
with $(1 + \Delta)^{-1} = 0.96 \cong 1$, we have consistency with perturbative evolution,
as expected. For sufficiently large $W^2$, or $10^{-3} > x > 10^{-5}$, we observe the expected
transition from unmodified pQCD evolution to hadronlike saturation.

We turn to the discussion of the gluon distribution function. Rewriting (\ref{eq:7.6}), and
substituting the gluon distribution from (\ref{eq:2.4}), we find
\bqa
&\frac{\frac{\partial F_2 (x,Q^2)}{\partial \ln Q^2}}{1+\Delta (\eta, \mu)} = F_L
\left( \frac{\xi_L}{\xi_2} x, Q^2 \right) \nonumber \\
&= \frac{\alpha_s (Q^2)}{9 \pi} R_{e^+e^-} G \left( \frac{x}{\xi_2}, Q^2 \right),
\label{eq:7.15}
\eqa
where $R_{e^+e^-} = 10/3$ for four flavors of quarks. The factor $(1+\Delta)^{-1}$ in
(\ref{eq:7.15}), for $\Delta \not= 0$ represents a violation of, or alternatively, a correction to the evolution equation for the gluon distribution with decreasing $\eta$ for
$\eta (W^2 = Q^2/x, Q^2) \lsim 5$.

The violation of the evolution with decreasing $Q^2$ may be seen directly in terms of the gluon distribution by showing the gluon distribution as a function of $x$ for e.g. $Q^2=100~{\rm GeV}^2$ and $Q^2=2~{\rm GeV}^2$ i.e. rewriting  (7.15) as  
\bqa
 G \left( {x}, Q^2 \right)=\frac{\frac{\partial F_2 (\xi_2{x},Q^2)}{\partial \ln Q^2}}{1+\Delta (\eta, \mu)} 
 \frac{9 \pi}{\alpha_s (Q^2) R_{e^+e^-}},
\label{eq:7.16}
\eqa
where $\Delta (\eta, \mu)$ is inserted from (7.6). In Fig.18 we show the gluon distribution according to (7.16). In addition, we show the gluon distribution under the ad hoc assumption  of  $\Delta (\eta, \mu){\equiv}0$ corresponding to validity of evolution in violation of the CDP result where  $\Delta (\eta, \mu) \neq 0$.
\begin{figure}[h]
\includegraphics[width=9cm]{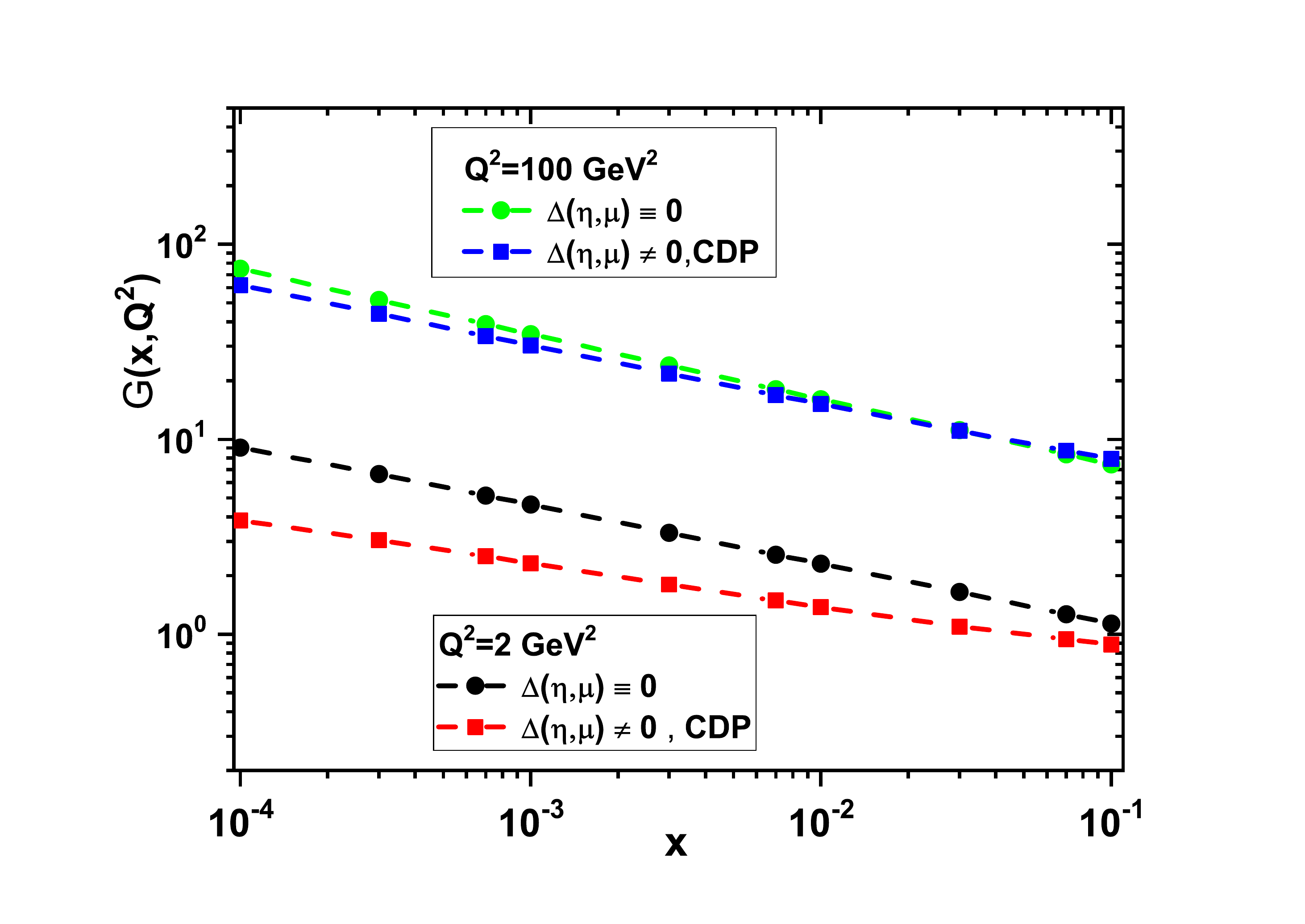}%
\caption{\label{Fig7}The gluon distribution $G (x,Q^2)$ deduced from
Eq.(7.16) for $\Delta(\eta,\mu){\neq}0$  and  $\Delta(\eta,\mu){\equiv}0$ .
}
\end{figure}

 Fig.18 confirms our previous  conclusion that the experimental data described by the CDP obey evolution at $Q^2$ large, $Q^2{\cong}100~\textrm{GeV}^2$, while evolution is violated at low $Q^2$, i.e., specifically for $Q^2{=}1.9~\textrm{GeV}^2$. Since the experimental data, described by the $\gamma^{*}g{\rightarrow}q\bar{q}$ interaction of the CDP, violate the evolution equation at low $Q^2$, the assumption of a universal validity of the evolution equation at low $Q^2$ is excluded. Relying on the universal absolute validity of the evolution equation at a low-$Q^2$ input scale in e.g. global fits to the experimental data is highly questionable.

The Ratio may be explicitly interpreted as a modification of the evolution equation (6.6).  Upon substitution of (7.10) equation (6.6) becomes 
\bqa
&{\frac{\partial F_2 (x,Q^2)}{\partial \ln Q^2}} = F_L
\left( \frac{\xi_L}{\xi_2} x, Q^2 \right){\times}Ratio\nonumber\\
&= F_L
\left( \frac{\xi_L}{\xi_2} x, Q^2 \right)\frac{1}{I^{(1)}_L \left( \frac{\xi_L}{\xi_2}\eta, 
\frac{\xi_L}{\xi_2} \mu \right)}  \nonumber \\
& \times  \left[ I_T^{(1)} \left( \frac{\eta}{\rho}, \frac{\mu}{\rho} \right)
+ I^{(1)}_L (\eta, \mu) \right. \nonumber \\
& \left. + \frac{\partial}{\partial \ln Q^2}
\left(I_T^{(1)} \left(\frac{\eta}{\rho}, \frac{\mu}{\rho} \right) +
I_L^{(1)} (\eta, \mu) \right) \right],
\label{eq:7.17}
\eqa
where the simple equations for $I_{L,T}^{(1)} $ in Appendix (A3) to (A5) are to be inserted. At small $\mu$, the CDP evolution equation (7.17)  becomes
\bqa
&\frac{\partial F_2 (x,Q^2)}{\partial \ln Q^2}{\simeq }F_L( \frac{\xi_L}{\xi_2} x, Q^2)\nonumber\\
&{\times}\bigg{\{}
\frac{I_{0}(\eta)-\frac{(1-C_{2})\eta}{1+4\eta}(2I_{0}(\eta)+\frac{1}{\eta})}{1-2\eta^{*}I_{0}(\eta^{*})}
\bigg{\}}
\eqa
where $I_{0}$ is defined in (A.5). The factor multiplying the longitudinal structure function continuously converges towards unity at large $Q^2$. We note that the CDP evolution equation (7.17), differs from an ad hoc modification of conventional evolution, since it is an analytically determined smooth extrapolation to low $Q^2$ consistent with DIS measurements.

\section{VIII. Conclusions.}
\renewcommand{\theequation}{\arabic{section}.\arabic{equation}}
\setcounter{section}{8}\setcounter{equation}{0}
 \bigskip
 
 Employing the pQCD relation between the longitudinal proton structure function and the gluon distribution function allows one to define a gluon distribution associated with the proton structure functions of the CDP at low $x$ and any $Q^2{\geq}0$. The CDP gluon distribution depends on $W^2=Q^2/x$ and $Q^2$. For $Q^2$ sufficient large ($10~\mathrm{GeV}^2{\lesssim}Q^2{\lesssim}100~\mathrm{GeV}^2$ at presently available energies)  the CDP gluon distribution, multiplied by $\alpha_{s}(Q^2)$, converges towards the asymptotic limit proportional  to the saturation scale $\Lambda^2_{\mathrm{sat}}(W^2){\sim}(\frac{W^2}{1\mathrm{GeV}^2})^{C_{2}=0.29}$.
 

 The deviation between the CDP gluon distribution and several
 published distributions, at low $Q^2$ in particular, is a consequence
 of the representation
 of the structure functions of DIS in the CDP. The CDP structure
 functions based on two-gluon exchange explicitly include large as
 well as small values of $Q^2$, including the $Q^2 \to 0$ limit.
 The associated CDP gluon distribution is valid for large as well
 as small $Q^2$. The usual determination of the gluon distribution
 however is not based on a very detailed justification of deducing
 the gluon distribution from the experimental data at a low-$Q^2$
 input scale. The CDP explicitly includes the $Q^2 \to 0$ structure
 functions, in distinction from the standard approach based on a
 low $Q^2$ input scale that lacks a very detailed justification
 for a low $Q^2$ input scale.

 Concerning evolution, the dependence of the logarithmic derivative of the proton
 structure function on $x$ and $Q^2$, we find a CDP evolution equation that agrees
 with the standard model evolution equation in the limit of large $Q^2$ at any fixed low $x$.
 
  At low $Q^2$, where photoabsorption cross section becomes hadronlike, the CDP
  evolution differs significantly from the standard one by an explicitly analytically
  given factor. Relying on a universal validity of the parton model and standard evolution
  at a low-$Q^2$ starting scale, such as $Q^2{\cong}1.9~ \mathrm{GeV}^2$, as frequently
  employed in global fits, is questionable and requires further investigations.

 
 
 
 

\begin{appendix}

\renewcommand{\theequation}{\Alph{section}.\arabic{equation}}
\setcounter{section}{0}
\setcounter{equation}{0}
\renewcommand{\thefigure}{\Alph{section}.\arabic{figure}}
\setcounter{section}{1}
\setcounter{figure}{0}
\section*{Appendix A}
 
The explicit expressions for the functions $I_{L,T}^{(1)} \left( \eta
(W^2,Q^2), \mu (W^2) \right)$ in (\ref{eq:3.2}) are given by:

\bqa
I_{L,T} (\eta (W^2,Q^2),\mu (W^2)) && = I^{(1)}_{L,T} (\eta (W^2,Q^2), \mu (W^2))\nonumber \\
&&  \times \left( 1+0 \left( \mu (W^2)\right) \right),
\label{eq:A.1}
\eqa
where $I^{(1)}_L (\eta (W^2,Q^2), \mu (W^2))~ {\rm and}~ I^{(1)}_T 
(\eta (W^2,Q^2), \mu (W^2))$
are given by
\bqa
&& I^{(1)}_L (\eta, \mu) = \frac{\eta - \mu}{\eta} \nonumber \\
&& \times \left( 1 - 
\frac{\eta}{\sqrt{1+4 (\eta - \mu)}} \right. \nonumber \\
&& \left.  \times \ln 
\frac{\eta (1+ \sqrt{1+4(\eta - \mu)})}{4 \mu-1-3\eta + 
\sqrt{(1+4(\eta - \mu))((1+\eta)^2 - 4 \mu)}} \right), \nonumber \\
&& I^{(1)}_T (\eta, \mu)  =  \frac{1}{2} \ln \frac{\eta-1 + 
\sqrt{(1+\eta)^2 - 4 \mu}}{2 \eta} \nonumber \\
&& - \frac{\eta - \mu}{\eta} + \frac{1+2(\eta - \mu)}
{2 \sqrt{1+4 (\eta - \mu)}}  \label{eq:A.2} \\
&&~~ \times  \ln \frac{\eta (1 + \sqrt{1+4 (\eta - \mu)})}
{4 \mu -1-3\eta + \sqrt{(1+4(\eta-\mu))((1+\eta)^2-4 \mu)}}.\nonumber 
\eqa
For $\mu (W^2) \ll 1$ (and $\eta (W^2,Q^2) \ge \mu (W^2)$) the functions
$I_L^{(1)} (\eta, \mu)$ and $I_T^{(1)} (\eta \mu)$ can be simplified to become
\be
I_L^{(1)} (\eta, \mu) = \frac{\eta - \mu}{\eta} \left( 1 - 2 \eta I_0 (\eta) \right)
\label{eq:A.3}
\ee
and
\be
I_T^{(1)} (\eta, \mu) = I_0 (\eta) - \frac{\eta - \mu}{\eta} \left( 1- 2 \eta
I_0 (\eta) \right),
\label{eq:A.4}
\ee
where
\be
I_0 (\eta) = \frac{1}{\sqrt{1 + 4 \eta}} \ln 
\frac{\sqrt{1 + 4 \eta} + 1}{\sqrt{1 + 4 \eta} - 1}.
\label{eq:A.5}
\ee
We also note that for the relevant range of $\eta \gg \mu$,
or $Q^2 \gg m^2_0$, we may put $\mu = 0$ in (\ref{eq:A.3}) and
(\ref{eq:A.4}).
\end{appendix}

\begin{appendix}


\renewcommand{\theequation}{\Alph{section}.\arabic{equation}}
\setcounter{section}{0}
\setcounter{equation}{0}
\renewcommand{\thefigure}{\Alph{section}.\arabic{figure}}
\setcounter{section}{1}
\setcounter{figure}{0}
\renewcommand{\thetable}{\Alph{section}.\arabic{table}}
\setcounter{section}{2}
\setcounter{table}{0}
\section*{Appendix B}

 
In this Appendix we show the coefficients of the structure functions
$F_2 (x,Q^2)$ given in (\ref{eq:4.1}) as taken from ref. \cite{Block}.
The coefficients in (\ref{eq:4.1}),
\bqa
A(Q^2) & = & a_0 + a_1 \ln \left(1 + \frac{Q^2}{\mu^2}\right) + a_2
\ln^2 \left( 1 + \frac{Q^2}{\mu^2} \right), \nonumber \\
B(Q^2) & = & b_0 + b_1 \ln \left(1 + \frac{Q^2}{\mu^2}\right) + b_2
\ln^2 \left( 1 + \frac{Q^2}{\mu^2} \right), \nonumber \\
C(Q^2) & = & c_0 + c_1 \ln \left(1 + \frac{Q^2}{\mu^2}\right), \nonumber \\
D(Q^2) & = & \frac{Q^2(Q^2 + \lambda M^2)}{(Q^2 + M^2)^2}, \nonumber
\eqa
depend on the fit parameters listed in Table II.
\begin{table}[h]
\caption{The effective parameters at low $x$ for $0.15 {\rm GeV}^2 <
Q^2 < 3000 {\rm GeV}^2$ provided by the following values. The fixed
parameters are defined by the Block-Halzen fit to the real photon-proton
cross sections as $M^2 = 0.753 \pm 0.068~ {\rm GeV}^2$, $\mu^2 = 2.82 \pm 0.290 ~{\rm GeV}^2$
and $c_0 = 0.255 \pm 0.016$.}
\begin{tabular}{|c|c|}
\hline
parameters & value \\
\hline
$a_0$ & $8.205 \times 10^{-4} \pm 4.62 \times 10^{-4}$\\
\hline
$a_1$ & $-5.148 \times 10^{-2} \pm 8.19 \times 10^{-3}$\\
\hline
$a_2$ & $-4.725 \times 10^{-3} \pm 1.01 \times 10^{-3}$\\
\hline \hline
$b_0$ & $2.217  \times 10^{-3} \pm 1.42 \times 10^{-4}$ \\
\hline
$b_1$ & $ 1.244 \times 10^{-2} \pm 8.56 \times 10^{-4}$ \\
\hline
$b_2$ & $ 5.958 \times 10^{-4} \pm 2.32 \times 10^{-4}$ \\
\hline\hline
$c_1$ & $1.475 \times 10^{-1} \pm 3.025 \times 10^{-2}$ \\
\hline
$n$ & $ 11.49 \pm 0.99$ \\
\hline
$\lambda$ & $ 2.430 \pm 0.153$ \\
\hline
$\chi^2$ (goodness of fit) & $0.95$ \\
\hline
\end{tabular}
\end{table}
\end{appendix}

\end{document}